\documentclass[10pt,amsfonts]{article}

\usepackage{amsmath}
\usepackage{amssymb}
\usepackage{bm}
\usepackage{graphicx}
\usepackage{xcolor}
\usepackage{caption}
\usepackage{subcaption}
\usepackage{appendix}
\usepackage{bigints}
\usepackage{yfonts}
\usepackage{bbold}
\usepackage[T1]{fontenc} 
\usepackage{amssymb}
\usepackage{bm}
\usepackage{graphicx}
\usepackage{xcolor}
\usepackage{authblk}
\usepackage{hyperref}
\usepackage{multirow} 
\usepackage{hhline}

\newcommand{\be}{\begin{equation}}
\newcommand{\ee}{\end{equation}}
\newcommand{\ba}{\begin{array}}
\newcommand{\ea}{\end{array}}
\newcommand{\bqa}{\begin{eqnarray}}
\newcommand{\eqa}{\end{eqnarray}}
\newcommand{\bea}{\begin{eqnarray}}
\newcommand{\eea}{\end{eqnarray}}

\begin{document}

\newtheorem{defi}{Definition}[section]
\newtheorem{lem}[defi]{Lemma}
\newtheorem{prop}[defi]{Proposition}
\newtheorem{theo}[defi]{Theorem}
\newtheorem{rem}[defi]{Remark}
\newtheorem{cor}[defi]{Corollary}

\title{Isospectral potentials with Dirac delta interaction: Constrained Spectra}

\author[1]{Kumar Abhinav \thanks{kumar.abh@mahidol.ac.th}}
\affil[1]{Centre for Theoretical Physics and Natural Philosophy, Nakhonsawan Studiorum for Advanced Studies, Mahidol University, Nakhonsawan, 60130, Thailand}
\author[2]{Biswanath Rath \thanks{biswanathrath10@gmail.com}}
\affil[2]{Department of Physics, Maharaja Sriram Chandra Bhanja Deo University, Takatpur, Baripada 757003, Odisha, India}
\author[3,4]{Prasanta K. Panigrahi \thanks{pprasanta@iiserkol.ac.in}}
\affil[3]{Department of Physical Sciences, Indian Institute of Science Education and Research Kolkata, Mohanpur 741246, West Bengal, India}
\affil[4]{CQST, Siksha O Anusandhan, Khandagiri, Bhubaneswar, 751030, Odisha, India}

\date{~~~}
\maketitle
\abstract{We study quantum potentials containing Dirac delta function (DDF) within its supersymmetric (isospectral) construction via a discontinuous superpotential, instead of being simply an added part. This construction comprises of two independently solvable sectors joined at the singular point by matching conditions imposed by the corresponding self-adjoint extension. Needing careful regularization, these boundary conditions impose independent algebraic constraints on each of the eigenstates, that restrict and truncate the spectrum. The number of surviving bound states, which are required to be parity-even, is thus fixed by the available system parameters rather than by the usual quantization rule. The specific cases of the Harmonic oscillator and the Rosen-Morse potential isospectrally infused with the DDF demonstrate such highly restrictive spectra, while the similar combination of the DDF with a Calogero-type singularity turns out to be largely incompatible as the regularization breaks down. Therefore, localized singularities are capable of controlling and engineering the discrete spectrum of quantum systems, with possible use in myriads of physical systems with very localized interactions.}

\section{Introduction}\label{S1}
Singular potentials model a number of physical systems despite the absence of measurably infinite interaction strengths in the real world. They serve as ideal limits of situations where the interactions become very strong over a seemingly negligible distance, such as near lattice points and atomic nuclei \cite{Belloni2014DeltaPotential}. The quantum mechanical treatment for singular potentials is characterized by the self-adjoint extension of the system \cite{GitmanTyutinVoronov2012}. In such cases, the corresponding wavefunctions can also be singular yet normalizable depending on boundary conditions at singularities. The particular case of the inverse-square (or Coulomb) potential requires renormalization based on the self-adjoint parameters to obtain bound states \cite{PhysRevA.89.022113} whereas square-integrability requires proper regularization \cite{Alhaidari2014}. 

\paragraph*{}A structurally simpler situation occurs when the potential takes the form of the Dirac delta function (DDF) \cite{griffiths2017introduction}. This is the simplest idealization of sharply localized interactions, and its distributive nature results in the characteristic integrability property. Moreover, the corresponding self-adjoint extension \cite{Albeverio2005Solvable} culminates in simple boundary conditions \cite{griffiths2017introduction}. In physical systems, the DDF can approximate potentials with a range much smaller than the characteristic length scales, such as the electron's wavelength in the medium or the lattice constant. A potential profile of the type $V(x)=\lambda\delta(x-x_0)$ is adequate for studying phenomena at relatively long wavelengths where the relevant physics is entirely captured by the strength parameter $\lambda$. As a consequence, the DDF adequately captures the scattering properties of arbitrary potentials \cite{Belloni2014DeltaPotential,Drukarev1978ZRP}, with the added benefit of ensuring exact solvability for such systems \cite{Belloni2014DeltaPotential}. 

\paragraph*{}Naturally, the DDF has been used to model various physical interactions, such as potentials due to sites in a crystal in the Kronig-Penney model \cite{10.1098/rspa.1931.0019}, quantum optical effects in mesoscopic systems \cite{BRANDES2005315}, in lattice QCD \cite{Bicudo2023} and many more \cite{Belloni2014DeltaPotential}. Furthermore, randomized distribution of DDFs models impurities and defects in many systems, such as irregularities in 1-dimensional lattices \cite{Gadella2021ContactInteractions,Belloni2014DeltaPotential}, impurity effects in Bose-Einstein condensates in an optical lattice \cite{PhysRevA.74.023614}, and impurities in discrete nonlinear systems \cite{Samiun2025ImpuritiesNLSE}. Such an aperiodic arrangement of point interactions represents an exactly solvable disordered system \cite{PhysRev.120.1175} that can support Anderson localization \cite{PhysRev.109.1492,Mott01041961} instead of a Bloch-like periodicity. As for the corresponding two-body scattering effect, enforced by energy–momentum conservation and permutation symmetry, the DDF interaction satisfies the Yang–Baxter equation and underlies the integrability of the delta-interacting Bose and Fermi gasses \cite{PhysRevLett.19.1312,PhysRev.168.1920}. It further represents a manifestation of confinement-induced resonances (CIRs) through scattering \cite{DUNJKO2011461}. 

\paragraph*{}The DDF possesses distinctly unique properties as an interaction. The bound and scattering states of the $N=2$ superextended DDF potential comprise a single algebraic supermultiplet under a hidden nonlinear $su(2\vert 2)$-type superunitary symmetry \cite{CORREA2008746}. At a fundamental level, in two dimensions, the Dirac delta potential renders a paradigmatic quantum-mechanical realization of the scale (conformal) anomaly via the dimensional transmutation of its classically dimensionless coupling \cite{Jackiw:216395,PhysRevD.68.125013}. In three dimensions, the universal unitary limit of two-body interactions is captured by the DDF interaction, which is central to cold-atom physics, particularly in the BCS–BEC crossover \cite{Pricoupenko_2007}. Furthermore, the pairwise delta interactions in the Calogero–Moser system generate an effective $1/r^2$ interaction upon hyperradial reduction \cite{PhysRevLett.82.463}. Consequently, in the case of the three-body interaction, regularizing the singular hyperradial equation with an extra boundary condition produces the discretely scale-invariant tower of Efimov states \cite{EFIMOV1973157,10.1119/1.3533428} that can be viewed as a renormalization-group limit cycle \cite{PhysRevB.69.094304}. In the Aharonov–Bohm scattering of particles with spin, the corresponding S-matrix depends on the self-adjoint extension due to the singular delta-function spin–flux (Zeeman) term \cite{PhysRevLett.64.503}. Furthermore, the bound spectrum of a Dirac fermion in Rindler spacetime, subject to genuine Dirac delta potentials, can directly represent the non-trivial zeros of the Riemann zeta function \cite{sym11040494,Kalauni_2023}, a construction structurally paralleled by the inverse-square potential \cite{DAS2019265}. 

\paragraph*{} Due to its absolute localization, the effect of the DDF potential manifests completely in terms of the modified boundary condition:
\begin{equation}
    \left.\frac{d\psi}{dx}\right\vert_{x=x_0}=\lambda\psi(x_0),\label{Eq00}
\end{equation}
of the eigenfunction \cite{Albeverio2005Solvable} at the location $x_0$ of the singularity. This simple condition completely encompasses the self-adjoint extension required by the singularity in the system. This simplicity prompts numerous pedagogical examples of pure DDF \cite{flugge2012practical,griffiths2017introduction} as well as DDF combined with other standard potential functions in quantum mechanics \cite{Belloni2014DeltaPotential}. It is seen that the unique yet simple boundary conditions imposed by the DDF contribution necessarily impact only those eigenfunctions of the remaining part of the potential that are even under reflection about the location of the DDF \cite{flugge2012practical,Belloni2014DeltaPotential}. Usually, the effect is a dip in the amplitude at the location of the DDF satisfying the discontinuity in the derivative of the eigenfunction at the location of the DDF. The few examples of potentials considered with the DDF were the infinite potential well and harmonic oscillator \cite{Belloni2014DeltaPotential}, which are symmetric about the DDF contribution at the origin. Naturally, the eigenstates `unperturbed' by the DDF contribution are alternatively even and odd, and only the even half of the spectrum is perturbed with a dip at the origin and a shift in energy.   

\paragraph*{} How the presence of DDF influences the isospectral structure of the system, given that the parity-based selective modification of the spectra is also of interest. Infinite potentials \cite{PhysRevD.32.1597} and inverse-square singularities \cite{JEVICKI198455,COOPER1995267} had been thought to break supersymmetry by influencing the superpartners to belong to different Hilbert spaces. This, however, was later found not to be the case by carefully regularizing such systems in order to approach the singularity \cite{DAS1999357}. A systematic treatment of generating various singular and non-singular superpotentials \cite{PANIGRAHI1993251} further led to a hierarchy of systems with both broken and unbroken supersymmetry. The case of the DDF singularity would be interesting since it manifests as simple boundary conditions, and the corresponding effect on the isospectral structure will be of interest. Given that the regularization or limiting approach for the DDF potential is quite straight-forward \cite{flugge2012practical}, its interplay with the supersymmetric structure of the overall system can be of practical interest.

\paragraph*{} In the present work, we explore the DDF singularity as a part of the isospectral structure by introducing a $\text{Sgn}(x)$ term in the superpotential. This is quite different from having an isospectral system in addition to a DDF singularity \cite{Belloni2014DeltaPotential} and is more akin to the isospectral treatment of $x^{-2}$-type potentials \cite{JEVICKI198455,COOPER1995267,DAS1999357,PANIGRAHI1993251}. Since the DDF is implemented essentially as a set of strict but simple boundary conditions \cite{griffiths2017introduction} stemming from the corresponding self-adjoint extension \cite{Albeverio2005Solvable}, its effect on the isospectral structure is of unique interest, unlike the other systems with a gradual singularity. It is found that the careful regularization treatment for the DDF \cite{flugge2012practical} leads to partial or complete disruption of the isospectral structure by imposing strong independent constraints, which is distinct from the systems where an isospectral system is just added to the DDF \cite{Belloni2014DeltaPotential}. The number of free parameters in the system is found to determine the allowed number of states subject to the said constraints. 

\paragraph*{} We start with a general view of the isospectral structure when the superpotential has a $\text{Sgn}(x)$ term in section \ref{S2}. It is found that the appearance of unavoidable singularities in the excited states naturally leads to a regularized system like that in Ref. \cite{flugge2012practical}, which is segmented. The simple system of DDF in a harmonic trap is considered in section \ref{S3} wherein the constraints force the spectrum to collapse. Section \ref{S4} deals with a Rosen-Morse type system with the DDF, revealing a finite number of states based on the number of available parameters of the system, although isospectrality is not present. Finally, we discuss the situation of isospectral systems where DDF is aided by the inverse-square contribution in section \ref{S5} before concluding in section \ref{S6} following some discussions.


\section{Isospectral treatment with DDF}\label{S2}
The presence of the DDF singularity in the potential obstructs a general solution of the Schr\"odinger equation that is valid at all locations. The natural approach to such situations is a regularization procedure that converges to the DDF singularity under suitable limits. On the other hand, an exactly solvable system must possess a supersymmetric (isospectral) structure ensuring its square integrability \cite{COOPER1995267}. As discussed before, in the case of the superposition of the DDF contribution $-\Omega\delta(x)$ on well-known solvable potentials \cite{Belloni2014DeltaPotential}, the arbitrary strength parameter $\Omega$ determines the distortion of the eigenfunctions at the location of the DDF (Eq. \ref{Eq00}), manifesting the self-adjoint extension. We, on the other hand, consider the systems where the DDF contribution is included within its isospectral structure of the potential so that the entire system is generated from a {\it single} superpotential $W(x)$. To have the DDF singularity in the potential, the superpotential necessarily contains a step discontinuity at the origin. As we will see, in addition to the obvious self-adjoint extension, this is a crucial feature that was absent in the previous solvable models with a n\'aive DDF contribution.

\paragraph*{}The singular potentials under consideration have the general form:
\begin{equation}
    V(x)=V_{\text{Reg.}}(x)-\Omega\delta(x),\quad \Omega>0,\label{E005}
\end{equation}
with $V_{\text{Reg.}}$ being well-behaved everywhere. The DDF contribution with a negative sign famously supports at least one bound state \cite{griffiths2017introduction}, even without any bounding contribution. For $x\neq 0$, $V(x)=V_{\text{Reg.}}(x)$ is locally solvable with eigenfunctions having a nice limit $\psi_{\text{Reg.}}(x\to 0^\pm)=\varphi_\pm$ approaching the DDF from both sides, where $\varphi_\pm$ are finite constants. The overall normalization of the system then demands, 
\begin{equation}
    \int_{-\infty}^{0^-}~dx~\left\vert\psi_{\text{Reg.}}(x<0)\right\vert^2+\int^{\infty}_{0^+}~dx~\left\vert\psi_{\text{Reg.}}(x>0)\right\vert^2=1.\label{Eq05}
\end{equation}
in addition to the continuity of the eigenfunction at the origin: $\varphi_+=\varphi_-=\varphi$, which serves as the solution at $x=0$. At the location of the DDF ($x=0$), the normalizable and well-behaved eigenfunctions satisfy the discontinuity condition \cite{griffiths2017introduction}:
\begin{equation}
    \left.\frac{d\psi}{dx}\right\vert_{x\to 0^-}^{x\to 0^+}=\lim_{\epsilon\to 0}\int_{-\epsilon}^\epsilon~dx~V(x)\psi(x)\equiv-\Omega\varphi.\label{Eq07}
\end{equation}
The above relation replaces the usual continuation of the derivative across $x=0$ and manifests the self-adjoint extension corresponding to the DDF \cite{Albeverio2005Solvable}.

\paragraph*{}We deliberately consider $V(x)=V_-(x)$ for some isospectral, shape-invariant pair $V_\pm(x)$. This is always possible for exactly solvable models with suitable parameterization, with the ground state $\psi(x)=\psi_0^-(x)$ directly derived from the corresponding superpotential. As a result, the self-adjoint extension in the case of the ground state specifically reduces to the expression for the superpotential $W(x)$ at the location of the DDF:
\begin{equation}
    \left.\frac{d}{dx}\psi_0^{-}(x)\right\vert_{x\to 0^-}^{x\to 0^+}\equiv-\left.W(x)\psi_0^{-}(x)\right\vert_{x\to 0^-}^{x\to 0^+}.\label{Eq08}
\end{equation} 
 In order to produce the DDF singularity in $V_-(x)$, the superpotential must have the general discontinuous form,
\begin{equation}
    W(x)=W_{\text{Reg.}}(x)+\omega(x),
\end{equation}
where $W_{\text{Reg.}}(x)$ is well-behaved and finite everywhere, whereas $\omega(x)$ behaves as $\omega(x\gtrless 0)=\omega_\pm,\,\,\omega_\pm\in\mathbb{R}$. Then Eq. \ref{Eq08} yields,
\begin{equation}
    \left.\frac{d}{dx}\psi_0^{-}(x)\right\vert_{x\to 0^-}^{x\to 0^+}\equiv-\left(\omega_+-\omega_-\right)\varphi,\label{Eq11}
\end{equation}
with the ground state being functionally distinct across $x=0$,
\begin{equation}
    \psi_0^{-}(x\gtrless0)=N_0\exp\left[-\int^x~dx~W(x)\right]=N_0\exp\left(-\omega_\pm x\right)\exp\left[-\int^x~dx~W_{\text{Reg.}}(x)\right],\label{EqN001}
\end{equation}
where $N_0$ is the normalization constant. Still, it is continuous and well-behaved everywhere as required. As for the potential itself, 
\begin{equation}
    V_-(x)=V_{\text{Reg.}}(x)-\frac{d}{dx}\omega(x),\quad V_{\text{Reg.}}(x)=W^2(x)-\frac{d}{dx}W_{\text{Reg.}}(x),
\end{equation}
not only are the functional forms for $x\gtrless 0$ different, but the finite part $V_{\text{Reg.}}(x)$ contains a contribution $2W_{\text{Reg.}}(x)\omega(x)$ that is discontinuous at $x=0$. The remaining part of $V_-(x)$ is the source of the DDF contribution,
\begin{eqnarray}
        \frac{d}{dx}\omega(x)=\left(\omega_+-\omega_-\right)\delta(x).
\end{eqnarray}
Therefore, the ground state satisfies Eq. \ref{Eq07} subject to the identification $\Omega=\omega_+-\omega_-$. 
This condition will also be satisfied {\it trivially} by any eigenstate that behaves as $\psi_n(0)=0$, which is the case for parity-odd eigenstates of parity-even potentials \cite{cohen1977quantum}. However, having two different superpotentials across the location of the DDF prohibits consistent parity-odd combination of individual eigenstates as they themselves will be discontinuous. This is because $x=0$ is not a symmetry point of either of the corresponding potentials, even if the overall potential distribution is symmetric. In the following examples we see that all states that survive in such a system are \textcolor{red}{parity-even}. For other situations such as even-parity combination of eigenstates or when parity is not a good quantum number, on the other hand, the excited states $\psi_{n>0}^-(x)$ will not satisfy Eq. \ref{Eq08}. However, since any eigenstate must satisfy Eq. \ref{Eq07}, this only implies that the strength of the DDF $\Omega$ will have a {\it different} expression than $\Omega=\omega_+-\omega_-$. Consequently, the system parameters will be constrained more and more for each allowed state through Eq. \ref{Eq07}. 

\paragraph*{}Moreover, the complete eigenfunctions can possess singularities of their own, given that the potential is singular. This is different from the cases where the DDF itself is not a part of the isospectral structure \cite{Belloni2014DeltaPotential,COOPER1995267}. Unlike the ground state, it is not always possible to infer the analytic structure of the excited state only from the knowledge of the superpotential. The {\it additional} presence of shape-invariance \cite{gendenshtein1983} rectifies the situation, since the superpartners are then related through $V_+(x;a_0)=V_-(x;a_1)+R(a_1)$. Here $a_0$ represents the set of initial parameters of the system, and $a_1$ represents a new set, with $R(a_1)$ being a constant function of parameters only. Consequently, for example, the 1st excited state of $V_-(x;a_0)$ can be written in terms of the ground state of $V_-(x;a_0)$ as,
\begin{equation}
    \psi_1^-(x;a_0)=\left(-\frac{d}{dx}+W(x;a_0)\right)\psi_0^-(x;a_1)=\left(W(x;a_1)+W(x;a_0)\right)\psi_0^-(x;a_1).\label{Eq16a}
\end{equation}
The shape-invariance cannot affect $\omega_\pm$ by construction. Notably, this eigenstate is itself discontinuous $x=0$, unlike the ground state due to its linear dependence on the superpotential. As a result, its derivative will be singular
exactly at $x=0$ and the LHS of Eq. \ref{Eq07} will not even be defined. This is a common feature of any excited state of a shape-invariant system with a discontinuous superpotential. The 2nd excited state has the form,
\begin{eqnarray}
    \psi_2^-(x;a_0)&=&\left(-\frac{d}{dx}+W(x;a_0)\right)\left(-\frac{d}{dx}+W(x;a_1)\right)\psi_0^-(x;a_2)\nonumber\\
    &=&\left[\left\{W(x;a_2)+W(x;a_0)\right)\left(W(x;a_2)+W(x;a_1)\right\}-\frac{d}{dx}W(x;a_2)-\frac{d}{dx}W(x;a_1)\right]\psi_0^-(x;a_2),\label{Eq016b}
\end{eqnarray}
which is itself singular at $x=0$ and so is its derivative. This pattern continues with higher states.

\paragraph{}Clearly, with the DDF integrated into the isospectral structure, a direct implementation of the derivative discontinuity condition (Eq. \ref{Eq07}) is not possible for excited states. However, the absolute locality of the DDF renders the superpotential and potentials well-behaved on both sides of $x=0$. Correspondingly, all the eigenfunctions are regular on either side. This motivates the regulating approach of having a finite potential around $x=0$ that approaches the DDF under some proper limit, with the regular part of the potential retained on both sides \cite{DAS1999357}. The simplest construction is a potential well of depth $V_0$ and width $2\varepsilon$ centered about $x=0$ such that for $V_0\to\infty,~\varepsilon\to 0$ we recover the DDF with $2\varepsilon V_0\to\Omega$ \cite{flugge2012practical}. As a result of avoiding the singularity at $x=0$, the usual continuity conditions for the eigenfunctions and their derivatives are regained at the interfaces $x=\pm\varepsilon$ of the potential distribution:
\begin{equation}
    V_-(x)=\begin{cases}
        \left(W_{\text{Reg.}}(x)+\omega_-\right)^2-\frac{d}{dx}W_{\text{Reg.}}(x), & x<-\varepsilon\\
        -V_0, & -\varepsilon\leq x\leq\varepsilon\\
        \left(W_{\text{Reg.}}(x)+\omega_+\right)^2-\frac{d}{dx}W_{\text{Reg.}}(x), & x>\varepsilon
    \end{cases}.\label{NE007}
\end{equation}
The two distinct and continuous superpotentials on either side of the well still represent the actual system with the DDF singularity now. Consequently, they correspond to individual eigenstates that are automatically the $x>\varepsilon$ and $x<-\varepsilon$ expressions of the {\it true} eigenstates, naturally approaching the complete ones as $V_0\to\infty,~\varepsilon\to 0$. This situation is clearly distinct from when the DDF is not included in the isospectral structure. In the latter case, parity-even systems like the infinite potential well \cite{flugge2012practical}, harmonic oscillator \cite{10.1119/1.9857}, and a few other systems \cite{Belloni2014DeltaPotential} with a DDF potential at the origin have been found to have all the eigenstates, with only the even states altered by the derivative discontinuity condition. Since the DDF was introduced only perturbatively \cite{Belloni2014DeltaPotential} to an isospectral potential defined by a {\it single} superpotential on both sides, the regularization approach and the direct implementation of Eq. \ref{Eq07} both led to the same spectra. In contrast, the present situation necessarily requires the regularization approach due to its singular isospectral nature. 

\paragraph*{} The present construction imposes compatibility of $-V_0$ with two different systems $W(x)=W_{\text{Reg.}}(x)+\omega_\pm$ at two different locations ($x=\pm\varepsilon$). As a result, {\it independent constraints} can be expected, unlike when both sides of the potential well have the same superpotential. Consequently, the individual systems are {\it not} necessarily parity-even about the origin, even if the complete system is. Since individual eigenstates on both sides may not have definite parity, every overall eigenstate will be affected by the boundary conditions that imbue the effect of the DDF. As we will see, this severely restricts the spectrum, since the two independent boundary conditions serve as independent constraints on the overall individual eigenfunctions. These features are explicated through three examples of the DDF isospectrally merged with physically relevant potentials with distinct structures. In the first, $V_-(x\gtrless0)$ are two shifted harmonic oscillators, while the second system consists of two different Rosen–Morse-type potentials. In the third one, $V_-(x\gtrless0)$ possesses Calogero-like profiles with their own $x^{-2}$-singularities in harmonic traps, depicting a competitive constraint structure.


\section{Isospectral combination of harmonic oscillator and DDF}\label{S3}
\begin{figure}[t]
    \centering
    \begin{subfigure}[b]{0.45\textwidth}
        \centering
        \includegraphics[width=\textwidth]{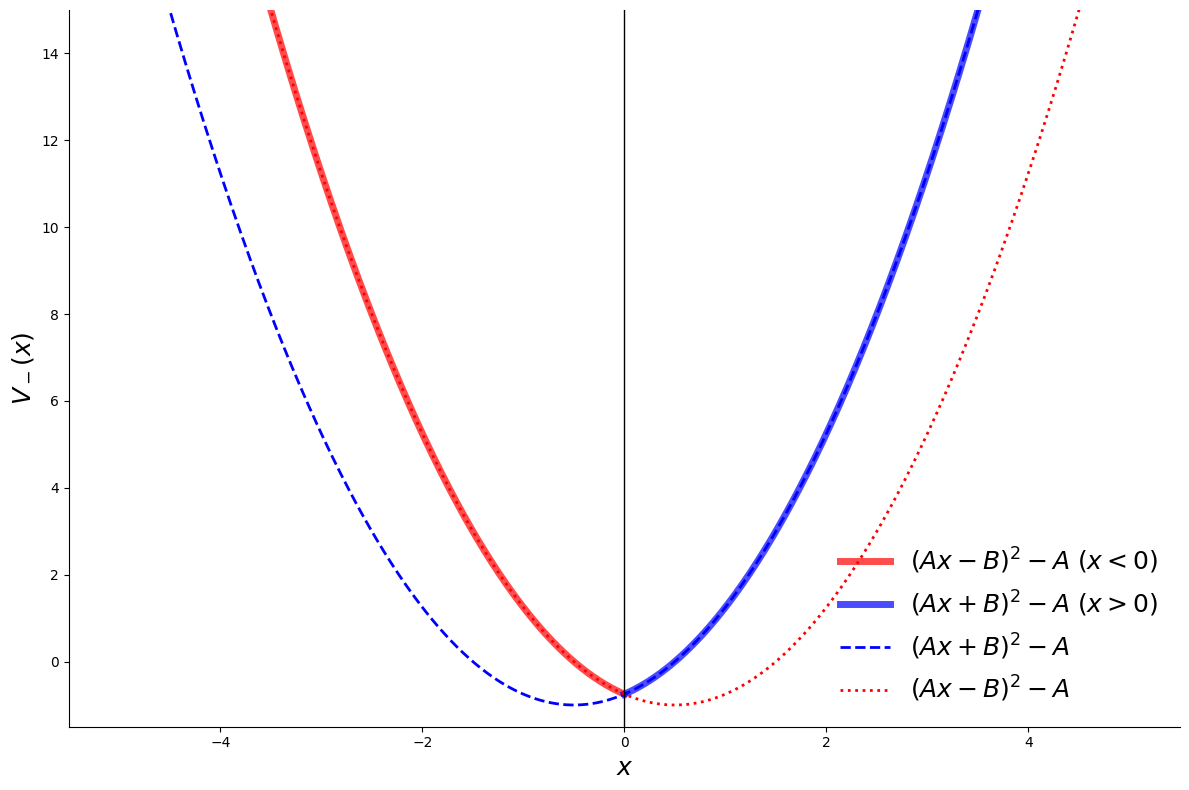}
        \caption{The regular potential (solid lines) for $B=0.5$. The dashed lines complete individual segments that are shifted oscillators centered at $x=\pm B/A$.}
    \label{F009}
    \end{subfigure}
    \hfill 
    \begin{subfigure}[b]{0.45\textwidth}
        \centering
        \includegraphics[width=\textwidth]{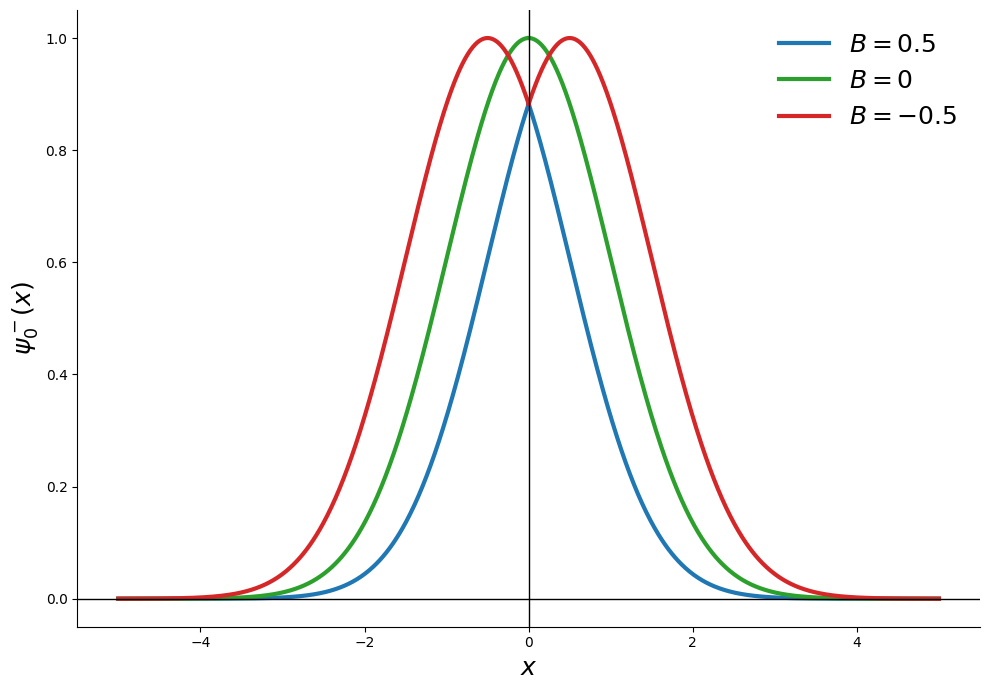}
        \caption{The ground state has a sharp peak (blue) or dip (red) for $B\gtrless0$. The Gaussian is recovered (green) for $B=0$.}
        \label{F009a}
    \end{subfigure}
    \caption{The simple harmonic oscillator with DDF: potential (regular part) and ground state. Here, $A=1$.}
    \label{Fig009}
\end{figure}

Being the most basic bound system in nature, the simple harmonic oscillator is also the most commonly used trap potential for many important systems characterized by various additional interactions \cite{RevModPhys.80.885}. The case of a DDF potential imposed on a simple harmonic oscillator potential has been discussed in detail \cite{RevModPhys.43.36}, with only parity-even eigenstates being affected by a dip at the location of the DDF and a shift in their energies. To include the DDF in the isospectral structure itself, we presently consider the simple superpotential and the corresponding superpartners as,
\begin{eqnarray}
    && W(x)=Ax+B~\text{Sgn}(x),\nonumber\\
    && V_\pm(x)=\left(Ax+B~\text{Sgn}(x)\right)^2\pm\left(A+2B\delta(x)\right).\label{ES02}
\end{eqnarray}
The regular part of this potential (Fig. \ref{F009}) partially combines two harmonic oscillators centered at {\it different locations} other than the origin. The limiting potential distribution of Eq. \ref{NE007} can now be invoked for this particular system as,
\begin{equation}
    V_-(x)=\begin{cases}
        (Ax+B)^2- A~~ \text{with}~W(x)=Ax+B & \text{for}~x>\varepsilon\\
        - V_0 & \text{for}~-\varepsilon\leq x\leq\varepsilon\\
(Ax-B)^2- A ~~ \text{with}~W(x)=Ax-B & \text{for}~x<-\varepsilon
\end{cases}.\label{ES03}
\end{equation}
The potentials in the regions $x>\varepsilon$ and $x<-\varepsilon$ are now even functions about the symmetry points $x=\mp B/A$ respectively. As a result, the alternate states for these individual systems also possess even and odd parities about the same points. Consequently, the individual odd-parity states do not vanish at the location of the DDF ($x=0$). As a result, the corresponding combined states are also affected by the discontinuity in their derivatives at $x=0$, just like the combination of individual even-parity states. 

\paragraph*{} Since the potential is locally a shifted harmonic oscillator subject to boundary conditions (Eq. \ref{ES03}) different from the usual ones, the general eigenfunction should be in terms of the parabolic cylinder functions \cite{AlHashimi2013} as,
\begin{equation}
    \psi_n^-(x\gtrless 0)=N_{n_\pm}^\pm D_{n_\pm}\left(\sqrt{\frac{2}{A}}y_\pm\right),\quad y_\pm=Ax\pm B,
\end{equation}
where the normalization constants $N_{n_\pm}^\pm$ account for the segmented nature of the potential. It is to be noted that the indices,
\begin{equation}
    n_\pm=\frac{1}{2}\left(\frac{E_\pm}{A}-1\mp 1\right),
\end{equation}
are {\it not} integers without boundary conditions of the harmonic oscillator, leading to suitably quantized energies $E_\pm$. Subsequently, subject to the potential distribution in Eq. \ref{ES03}, the distribution of the eigenfunctions of this system takes the form,
\begin{equation}
    \psi_n^-(x)=\begin{cases}
        \varphi_I(x)=N_{n_-}^-D_{n_-}\left(\sqrt{\frac{2}{A}}y_-\right) & \text{for}~x<-\varepsilon\\
\varphi_{II}(x)=C\cos(px)+D\sin(px) & \text{for}~-\varepsilon\leq x\leq\varepsilon\\
\varphi_{III}(x)=N_{n_+}^+D_{n_+}\left(\sqrt{\frac{2}{A}}y_+\right) & \text{for}~x>\varepsilon
    \end{cases},\label{ES05a}
\end{equation}
where $C,~D$ are the respective normalization constants and $p=\sqrt{V_0+E}\in\mathbb{R}$ is the momentum of the standing wave (bound state) inside the well. 

\paragraph*{}The implementation of the boundary conditions at $x=\pm\varepsilon$ can now follow as usual. However, we take advantage of the symmetry of the system to simplify the following calculations. From Fig. \ref{F009} the overall potential is even-parity. Equivalently, the overall Hamiltonian has to commute with the parity operator, and thus all its eigenstates in Eq. \ref{ES05a} have to posses odd or even parity. Overlooking the limiting barrier\footnote{Since the DDF does not hamper the continuity of the eigenfunctions.}, the complete eigenfunction has the form,
\begin{equation}
    \psi_{n_+,n_-}(x)=\Theta(x)N_{n_+}^+D_{n_+}\left(\sqrt{\frac{2}{A}}y_+\right)+\Theta(-x)N_{n_-}^-D_{n_-}\left(\sqrt{\frac{2}{A}}y_-\right),
\end{equation}
with the Heaviside step function $\Theta(x)$. Since $\hat{P}\psi_{n_+,n_-}(x)=\psi_{n_+,n_-}(-x)\equiv\pm\psi_{n_+,n_-}(x)$ is a must, the two parts of the eigenfunction are constrained to follow,
\begin{equation}
    N_{n_\pm}^\pm D_{n\pm}\left(-\sqrt{\frac{2}{A}}y_\mp\right)=\sigma N^\mp_{n_\mp}D_{n_\mp}\left(\sqrt{\frac{2}{A}}y_\mp\right),\quad\sigma=\pm 1.
\end{equation}
Moreover, since the parabolic cylinder functions $D_n(x)$ are linearly independent for different arguments $n$, the only way the above condition is met is when $N^+_{n_+}=\sigma N^-_{n_+}$ and more importantly, $n_+=n_-\equiv n\in\mathbb{Z}_{\geq 0}$. The later condition reduces the parabolic cylinder functions to the familiar Hermite polynomials enveloped by a gaussian:
\begin{equation}
    D_n(x)=2^{-n/2}H_n\left(\frac{x}{\sqrt{2}}\right)\exp\left(-\frac{x^2}{4}\right), \quad n\in\mathbb{Z}_{\geq 0}.\label{HerPol}
\end{equation}
Therefore, the overall eigenfunctions of the present system take the form\footnote{The factor $2^{-n/2}$ is absorbed into the normalization constant for brevity.},
\begin{equation}
    \psi_n^-(x\gtrless 0)=N_n^\pm H_n\left(\frac{y_\pm}{\sqrt{A}}\right)\exp\left(-\frac{y_\pm^2}{2A}\right).\label{ES04}
\end{equation}
 As a result, the well-known shape-invariant structure underlying these eigenfunctions \cite{COOPER1995267} will not be necessary for the following treatment, and the distribution of the eigenfunction of the limiting system now takes a simpler form:
\begin{equation}
    \psi_n^-(x)=\begin{cases}
        \varphi_I(x)=N_n^-H_n\left(\frac{y_-}{\sqrt{A}}\right)\exp\left(-\frac{y_-^2}{2A}\right) & \text{for}~x<-\varepsilon\\
\varphi_{II}(x)=C\cos(px)+D\sin(px) & \text{for}~-\varepsilon\leq x\leq\varepsilon\\
\varphi_{III}(x)=N_n^+H_n\left(\frac{y_+}{\sqrt{A}}\right)\exp\left(-\frac{y_+^2}{2A}\right) & \text{for}~x>\varepsilon
    \end{cases}.\label{ES05}
\end{equation}
which can now be subjected to the boundary conditions\footnote{The reduction of $D_n(x)$ to yield Hermite polynomials is also obtained from these {\it same} boundary conditions, as shown in Appendix \ref{A00}.}. The equality of the eigenfunction and its derivative at $x=-\varepsilon$ leads to the expressions,
\begin{eqnarray}
    &&C=\frac{N^-_n}{p}\exp\left(-\frac{\theta^2}{2A}\right)\left[p\cos(p\varepsilon)H_n\left(-\frac{\theta}{\sqrt{A}}\right)+\sin(p\varepsilon)\left\{2n\sqrt{A}H_{n-1}\left(-\frac{\theta}{\sqrt{A}}\right)+\theta H_n\left(-\frac{\theta}{\sqrt{A}}\right)\right\}\right],\nonumber\\
    &&D=\frac{N^-_n}{p}\exp\left(-\frac{\theta^2}{2A}\right)\left[-p\sin(p\varepsilon)H_n\left(-\frac{\theta}{\sqrt{A}}\right)+\cos(p\varepsilon)\left\{2n\sqrt{A}H_{n-1}\left(-\frac{\theta}{\sqrt{A}}\right)+\theta H_n\left(-\frac{\theta}{\sqrt{A}}\right)\right\}\right],\label{ES06a}
\end{eqnarray}
with $\theta=A\varepsilon+B$, whereas the corresponding boundary conditions at $x=\varepsilon$ yield,
\begin{eqnarray}
    &&C=\frac{N^+_n}{p}\exp\left(-\frac{\theta^2}{2A}\right)\left[p\cos(p\varepsilon)H_n\left(\frac{\theta}{\sqrt{A}}\right)-\sin(p\varepsilon)\left\{2n\sqrt{A}H_{n-1}\left(\frac{\theta}{\sqrt{A}}\right)-\theta H_n\left(\frac{\theta}{\sqrt{A}}\right)\right\}\right],\nonumber\\
    &&D=\frac{N^+_n}{p}\exp\left(-\frac{\theta^2}{2A}\right)\left[p\sin(p\varepsilon)H_n\left(\frac{\theta}{\sqrt{A}}\right)+\cos(p\varepsilon)\left\{2n\sqrt{A}H_{n-1}\left(\frac{\theta}{\sqrt{A}}\right)-\theta H_n\left(\frac{\theta}{\sqrt{A}}\right)\right\}\right].\label{ES06b}
\end{eqnarray}
Herein we had considered that the Hermite polynomials follow the Appell sequence: $\frac{d}{dx}H_n(x)=2nH_{n-1}(x)$. On equating the values of $C$ and $D$ from Eqs. \ref{ES06a} and \ref{ES06b} respectively by using the relation $H_n(-x)=(-1)^nH_n(x)$, we get,
\begin{equation}
   \text{either}\quad N_n^+=(-1)^nN_n^-:=(-1)^nN_n\quad\text{or}\quad N_n^+=(-1)^{n+1}N_n^-:=(-1)^{n+1}N_n,\label{ES06c}
\end{equation}
both of which cannot be true simultaneously. Thus, we have two cases at hand:
\begin{itemize}
    \item[{\bf Case 1}:] Choosing $N_n^+=(-1)^nN_n^-:=(-1)^nN_n$ naturally implies $D=0$, eventually leading to the condition,
    \begin{equation}
    p\tan(p\varepsilon)H_n\left(\frac{\theta}{\sqrt{A}}\right)=\theta H_n\left(\frac{\theta}{\sqrt{A}}\right)-2n\sqrt{A}H_{n-1}\left(\frac{\theta}{\sqrt{A}}\right).
\end{equation}
The system must respect this {\it parametric constraint} as a cost of compatibility among the boundary conditions at $x=\pm\varepsilon$. As a consequence, the surviving amplitude within the potential well has the form, 
\begin{equation}
    C=(-1)^nN_n\sec(p\varepsilon)H_n\left(\frac{\theta}{\sqrt{A}}\right)\exp\left(-\frac{\theta^2}{2A}\right).
\end{equation}
 \item[{\bf Case 2}:] The other choice $N_n^+=(-1)^{n+1}N_n^-:=(-1)^{n+1}N_n$ implied $C=0$ instead, leading to the condition,
 \begin{equation}
    p\cot(p\varepsilon)H_n\left(\frac{\theta}{\sqrt{A}}\right)=-\theta H_n\left(\frac{\theta}{\sqrt{A}}\right)+2n\sqrt{A}H_{n-1}\left(\frac{\theta}{\sqrt{A}}\right),
\end{equation}
which eventually yields the surviving amplitude in the well as,
\begin{equation}
     D=(-1)^{n+1}N_n\csc(p\varepsilon)H_n\left(\frac{\theta}{\sqrt{A}}\right)\exp\left(-\frac{\theta^2}{2A}\right).
\end{equation}
\end{itemize}
The energy quantum number $n$ decides the overall parity of the combined state in both cases since $H_n(x)$ carries a parity index of $(-1)^n$. 

\paragraph*{} The actual system with the DDF at $x=0$ is obtained through the limit $\varepsilon\to 0$ and $V_0\to\infty\Rightarrow p\to\infty$ such that $p\varepsilon\to 0$ but $p^2\varepsilon\to\Omega/2$ \cite{flugge2012practical}. Then in Case 1 the constraint condition reduces to,
\begin{equation}
    \boxed{\frac{\Omega}{2}H_n\left(\frac{B}{\sqrt{A}}\right)=B H_n\left(\frac{B}{\sqrt{A}}\right)-2n\sqrt{A}H_{n-1}\left(\frac{B}{\sqrt{A}}\right)},\label{ES11}
\end{equation}
in addition to the amplitude,
\begin{equation}
    C=(-1)^nN_nH_n\left(\frac{B}{\sqrt{A}}\right)\exp\left(-\frac{B^2}{2A}\right),\label{ES11a}
\end{equation}
which now represents the value of the combined eigenfunction at $x=0$. Since it is not a local extremum of either of the partial eigenfunctions in Eq. \ref{ES05} we have a discontinuous derivative at the $x=0$ as required. On the other hand, Case 2 yields the limiting expression of the constraint condition,
\begin{equation}
    \frac{1}{\varepsilon}H_n\left(\frac{B}{\sqrt{A}}\right)=BH_n\left(\frac{B}{\sqrt{A}}\right)-2n\sqrt{A}H_{n-1}\left(\frac{B}{\sqrt{A}}\right),\label{ES11A}
\end{equation}
requiring $B=\infty$ as $\varepsilon\to 0$ which cannot be valid. 

\paragraph{Ground state:} For the ground state ($n=0$), the constraint condition in Eq. \ref{ES11} of Case 1 reduces under the DDF limit to $ \boxed{\Omega=2B}$, representing a potential $-2B\delta(x)$ at the origin as required, in addition to $N_0^-=N_0^+=N_0$. This is the parity-even combination with the amplitude,
\begin{equation}
    C=N_0\exp\left(-\frac{B^2}{2A}\right),
\end{equation}
at the origin, which is either a peak or a dip depending on the sign of $B$, as plotted in Fig. \ref{F009a}. 

\paragraph*{} Note that this allowed ground state is identical to the one that can be directly obtained from Eq. \ref{EqN001} with the overall discontinuous superpotential. Thus, the strengths of the corresponding DDF singularities have to be identical as Eq. \ref{Eq08} is valid only for the ground state. Since shape-invariance exists for the segmented systems only, we will get $\Omega\neq 2B$ for the excited states.
Since the individual ground states from the segments are parity-even about $x=\mp B/A$, for $B/A>0$ the two parts combine into a peak at $x=0$. This is a recurring property of the ground state of such systems in general, as we will see. It is simple to check that the derivatives of the ground states across $x=0$ satisfy Eq. \ref{Eq07} as, 
\begin{eqnarray}
    &&\left.\frac{d\varphi_{III}}{dx}\right\vert_{x\to 0^+}-\left.\frac{d\varphi_I}{dx}\right\vert_{x\to 0^-}=-2BN_0\exp\left(-B^2/2A\right),\nonumber\\
    && \int_{-\varepsilon}^\varepsilon V(x)\psi_0^-(x)dx=-2BC=-2BN_0\exp\left(-B^2/2A\right).
\end{eqnarray}

\paragraph{1st excited state:}For $n=1$, Case 1 corresponds to $N_1^-=-N_1^+=N_1$, rendering a parity-even combination of parity-odd first excited states of individual harmonic oscillators. Since the symmetry points of the latter are not $x=0$, this is the only way the overall eigenfunction can be continuous there. Then Eq. \ref{ES11} reduces to,
\begin{equation}
    \boxed{\frac{\Omega}{2}=B-\frac{A}{B}}\quad\text{with}~~ C=-2N_1\frac{B}{\sqrt{A}}\exp\left(-\frac{B^2}{2A}\right),\label{ES18A}
\end{equation}
yielding an even combination peaked at $x=0$. As inferred, the present constraint condition is different from that for the ground state ($\Omega=2B$). The simultaneous validity of both imposes the severe restriction $A=0$ that reduces both eigenstates to the only bound state \cite{griffiths2017introduction} of the pure DDF potential. \footnote{This behavior is not apparent from Eq. \ref{ES05}, since the transition to continuum from any bound system is non-analytic. As a better indicator, the symmetry points of the individual oscillators become $x=\mp B/A\to\mp\infty$ for $A=0$.}. Therefore, a non-trivial first excited state is ruled out.

\paragraph{2nd excited state:} The situation for $n=2$ corresponds to individual states being parity-even ($N_2^-=N_2^+=N_2$) about $x=\mp B/A$. Under the DDF limit,
\begin{equation}
    \boxed{\frac{\Omega}{2}=B\frac{2B^2-5A}{2B^2-A}},\quad\text{with}~~ C=2\frac{N_2}{A}\left(2B^2-A\right)\exp\left(-\frac{B^2}{2A}\right),\label{ES26A}
\end{equation}
which is consistent with the ground state only if $A=0$. Therefore, again, a non-trivial excited state cannot survive.

\paragraph*{}It can be inferred that all the eigenstates either collapse to a single one peaked at the origin with no harmonic trap ($A=0$) or only the ground state survives in the Harmonic trap. In either case, the isospectral structure does not exist. This is entirely different from a harmonic trap with an added DDF \cite{RevModPhys.43.36}, clearly demonstrating the toll of having two different systems on either side. Since the harmonic oscillator is self-isospectral, the spectrum of $V_+(x)$ will be empty since its eigenfunctions are generated from the excited states of $V_-(x)$. However, in the later examples, we will observe that the lack of symmetry in individual systems might partially salvage the spectrum. 


\section{Rosen-Morse potential integrated to a DDF singularity}\label{S4}
In the preceding example, individual eigenstates of each segment had parity properties that mismatched with the symmetry of the overall system ({\it e. g.}, the location of the DDF), allowing only a single state. We now look into a system with two segments devoid of any such individual symmetries, isospectrally combined with the DDF singularity to form an overall parity-even potential distribution. The Rosen-Morse-type potentials, introduced to explain anharmonic vibrations in polyatomic molecules \cite{RosenMorse1932}, provide improved modeling of diatomic-molecular \cite{PhysRevA.86.062510,Abu-Shady2023} and dimer \cite{ONATE2021103961} spectra. They have been further utilized in relation to quark deconfinement \cite{doi:10.1142/S0219887824501913}, relativistic bound-states \cite{Oyewumi2010}, quantum wells \cite{condmat8040086} and quantum dots \cite{Ungan_2020}. Despite being devoid of any reflection symmetry, it is a non-trivial system displaying translational shape-invariance \cite{gendenshtein1983,COOPER1995267} with exactly solvable generalizations \cite{Bengherabi2025}, and therefore is suitable for the desired isospectral combination with the DDF singularity. This is achieved through the following discontinuous superpotential,
\begin{equation}
    W(x)=A\tanh(\alpha x)+B~\text{Sgn}(x),\label{E001}
\end{equation}
that leads to an isospectral pair of potentials,
\begin{equation}
    V_\pm(x)=A^2+B^2-A\left(A\mp\alpha\right)\text{sech}^2(\alpha x)+2AB\tanh(\alpha x)\text{Sgn}(x)\pm 2B\delta(x).\label{E002}
\end{equation}
The overall ground state for $V_-(x)$ can be immediately obtained as,
\begin{equation}
    \psi_0^-(x)=N_0^-\exp\left[-\int^x~W(x)dx\right]\equiv N_0^-\exp(-B\vert x\vert)\text{sech}^{A/\alpha}(\alpha x),\label{NE001}
\end{equation}
displays the usual discontinuous derivative at $x=0$. 

\begin{figure}[t]
    \centering
    \begin{subfigure}[b]{0.45\textwidth}
        \centering
        \includegraphics[width=\textwidth]{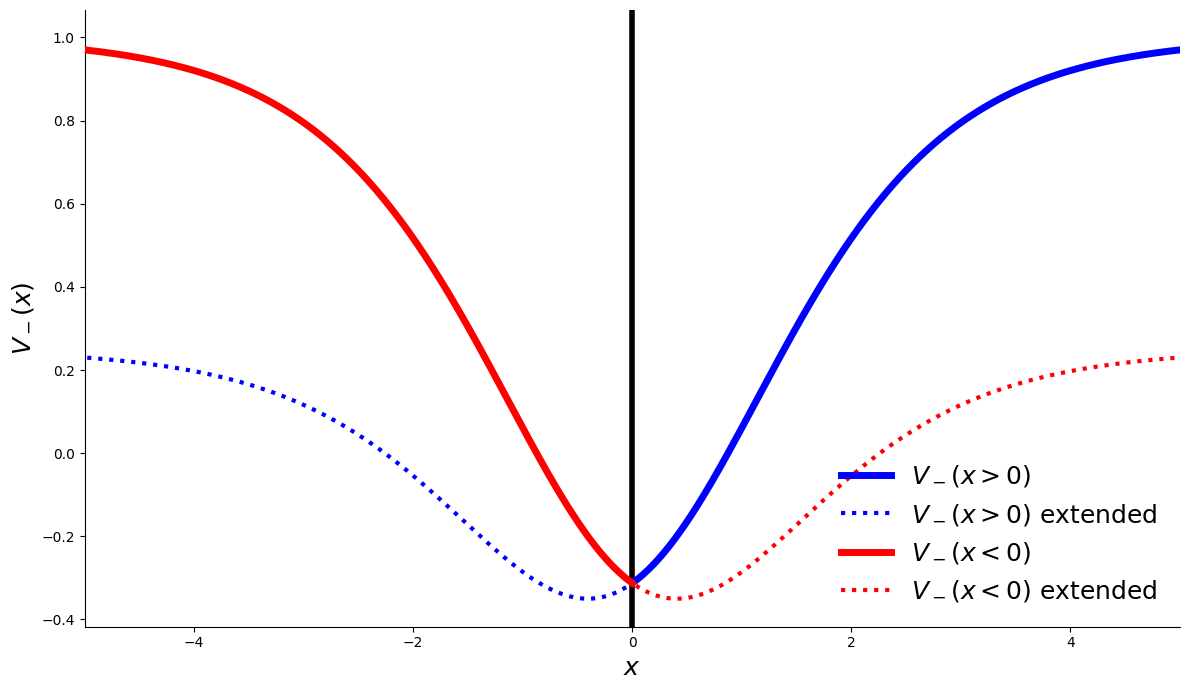}
        \caption{$V_-(x)$ (solid) combines two distinct potentials for $x\gtrless 0$ (blue/red), which lack definite parity (dotted extensions).}
    \label{F001}
    \end{subfigure}
    \hfill
    \begin{subfigure}[b]{0.45\textwidth}
        \centering
        \includegraphics[width=\textwidth]{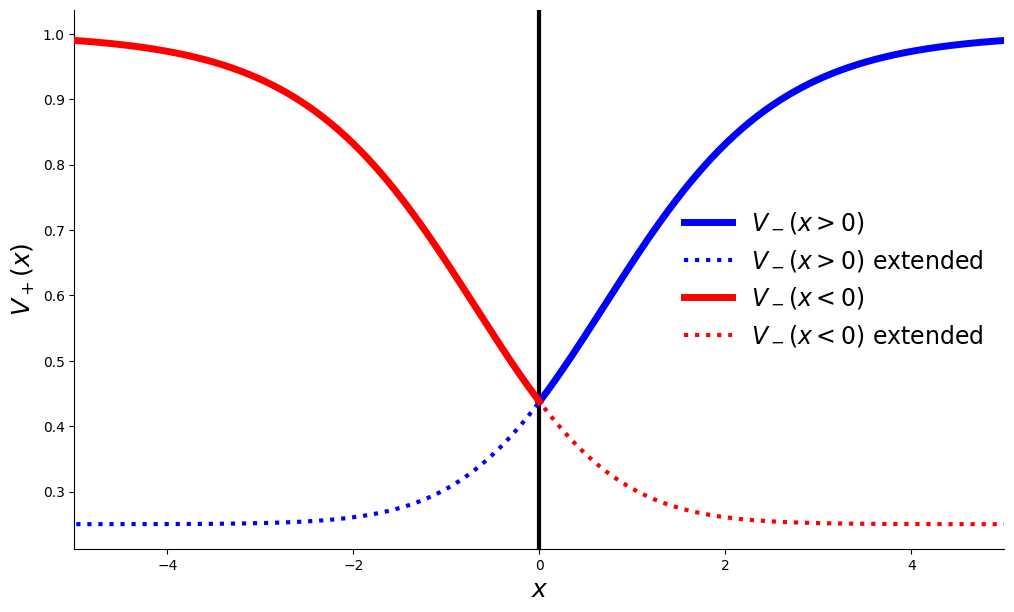}
        \caption{The superpartner $V_+(x)$ (solid) with constituents that are unbounded (dotted extensions). They are individually parity-odd, but not about $x=0$.}
        \label{F004}
    \end{subfigure}
    \caption{Combined Rosen-Morse-type potentials with parameter values $B=0.25$, $A=0.75$ and $\alpha=0.5$. The DDF contribution is represented by a vertical black line at $x=0$.}
    \label{Fig001}
\end{figure}

\paragraph*{} Across the DDF, the system resolves into two different isospectral potentials, 
\begin{eqnarray}
    &&W(x\gtrless0)=A\tanh(\alpha x)\pm B,\nonumber\\
    && V_+(x\gtrless0)=A^2+B^2-A\left(A-\alpha\right)\text{sech}^2(\alpha x)\pm 2AB\tanh(\alpha x),\nonumber\\
    && V_-(x\gtrless0)=A^2+B^2-A\left(A+\alpha\right)\text{sech}^2(\alpha x)\pm 2AB\tanh(\alpha x).\label{E003}
\end{eqnarray}
Both of them are additionally shape-invariant under the parameterizations $A_n=A-n\alpha,~B_n=AB/A_n$,
leading to the spectra $E_n=A^2+B^2-A_n^2-B_n^2$. This further enables the usual algebraic determination of the eigenfunctions starting from the ground state of Eq. \ref{NE001}.
The superpartners are plotted\footnote{The choice of parameter values conforms to the extensive constraint structure of the system that will emerge from the excited eigenfunctions.} in Fig. \ref{Fig001}. The combined potential $V_-(x)$ is indeed symmetric about the DDF location, although its individual constituents do not have any definite parity. On the other hand, $V_+(x)$ is a bounded system comprising two unbounded segments (Fig. \ref{F004}).

\subsection{The eigenstates of $V_-(x)$}
In order to utilize the isospectrality of the system, we start by regulating $V_-(x)$ as the distribution,
\begin{equation}
    V_-(x)=\begin{cases}
A^2+B^2-A\left(A+\alpha\right)\text{sech}^2(\alpha x)- 2AB\tanh(\alpha x) & \text{for}~ x<-\varepsilon, \\
-V_0 & \text{for}~ -\varepsilon\leq x\leq\varepsilon, \\
A^2+B^2-A\left(A+\alpha\right)\text{sech}^2(\alpha x)+ 2AB\tanh(\alpha x) & \text{for}~ \varepsilon<x.
\end{cases}\label{E018}
\end{equation}
Unlike the previous example, since no closed form expressions for the general eigenstate of these individual systems are available, we utilize their inherent shape-invariance to successively construct excited states from the ground states that obey Eq. \ref{NE001} on either side of $x=0$. Then the boundary conditions can be applied to each eigenstate individually in order to arrive at the DDF limit in each case. 

\subsubsection{Ground state}
\begin{figure}[t]
    \centering
    \begin{subfigure}[b]{0.45\textwidth}
        \centering
        \includegraphics[width=\textwidth]{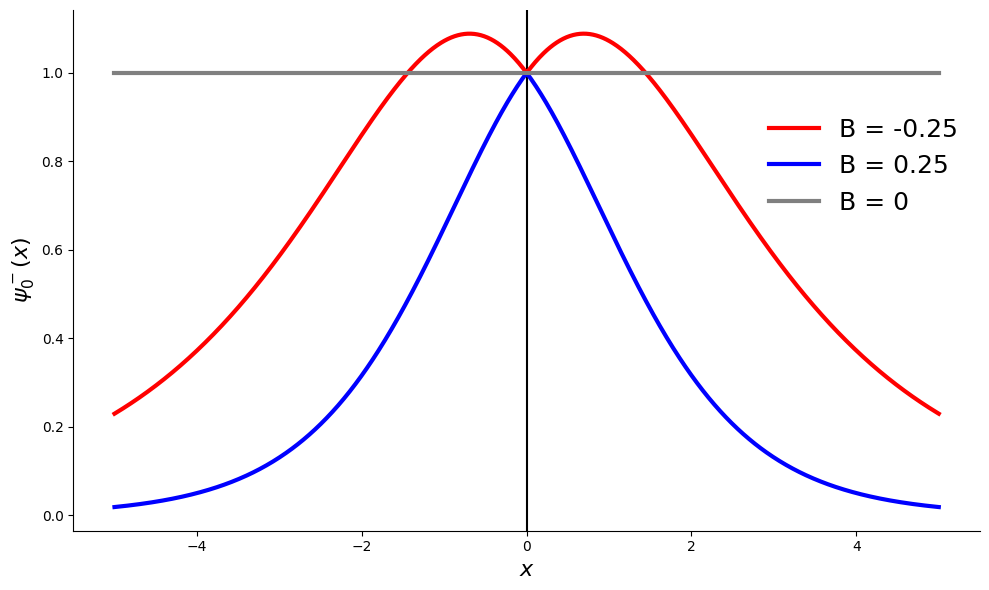}
        \caption{The ground state with $\alpha=2B$, $A=3B$. It vanishes for $B=0$ with the DDF.}
    \label{F002a}
    \end{subfigure}
    \hfill 
    \begin{subfigure}[b]{0.45\textwidth}
        \centering
        \includegraphics[width=\textwidth]{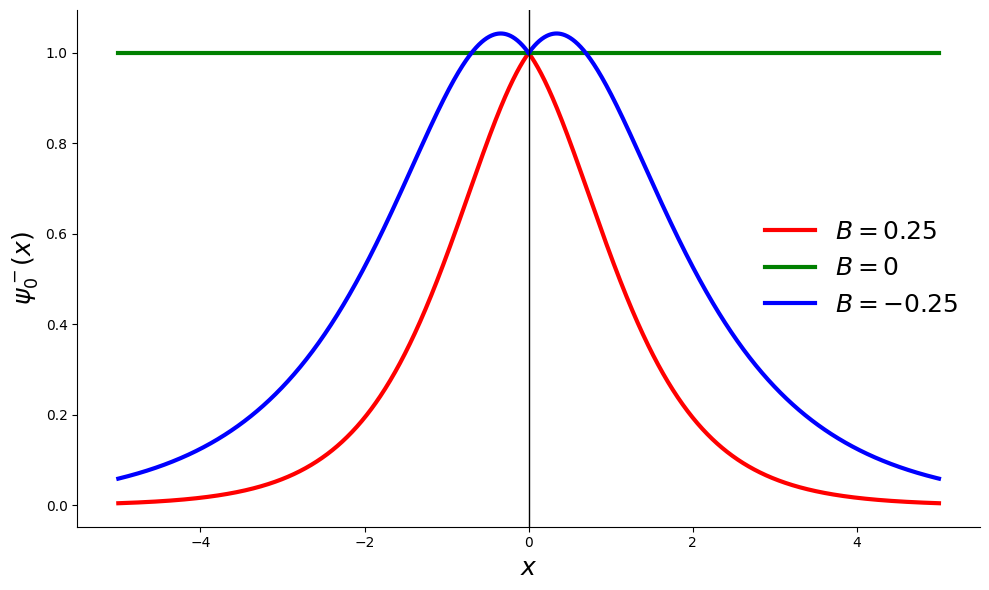}
        \caption{The ground state with $\alpha=-3B$, $A=-4B$. It vanishes for $B=0$ with the DDF.}
        \label{F002b}
    \end{subfigure}
    \caption{The ground state for the Rosen-Morse-type potential corresponding to two sets of parameters consistent with overall constraints (up to the second excited states).}
    \label{Fig002}
\end{figure}

The distribution of the combined ground state of $V_-(x)$ emerges from the superpotentials in Eq. \ref{E003} as,
\begin{equation}
    \psi_0^-(x)=
    \begin{cases}
        \varphi_I(x)=N_0^-\exp(Bx)~\text{sech}^{A/\alpha}(\alpha x)  & \text{for}~ x<-\varepsilon, \\
        \varphi_{II}(x)=C\cos(px)+D\sin(px) & \text{for}~ -\varepsilon\leq x\leq\varepsilon, \\
       \varphi_{III}(x)=N_0^+\exp(-Bx)~\text{sech}^{A/\alpha}(\alpha x)  & \text{for}~ \varepsilon<x.
    \end{cases}\label{E019}
\end{equation}
where $N_0^\pm, C, D$ are the respective normalization constants, and the potential well contains a standing wave of momentum $p=\sqrt{V_0+E}\in\mathbb{R}$. The boundary conditions at $x=\pm\varepsilon$ are applied to the above wavefunctions in Appendix \ref{A0}. The DDF limit is survived by the symmetric case with $N_0^-=N_0^+=N_0$ having $C=N_0$ and $D=0$, as depicted in Fig. \ref{Fig002}. Clearly, there is a discontinuity in the derivative at the origin. As before, the sign of strength $B$ of the singularity determines whether the ground state has a peak or a dip at the origin. Similar to the previous case, the DDF contribution affects the same constraint $\boxed{\Omega=2B}$ for the ground state, justifying Eq. \ref{Eq08}. Subsequently, the derivative discontinuity condition (self-adjointness) is validated as,
\begin{eqnarray}
    &&\left.\frac{d\varphi_{III}}{dx}\right\vert_{x\to 0^+}-\left.\frac{d\varphi_I}{dx}\right\vert_{x\to 0^-}=-2BN_0,\nonumber\\
    && \int_{-\varepsilon}^\varepsilon V(x)\psi_0^-(x)dx=-2BC=-2BN_0.
\end{eqnarray}

\subsubsection{First excited state}
\begin{figure}[t]
    \centering
    \begin{subfigure}[b]{0.45\textwidth}
        \centering
        \includegraphics[width=\textwidth]{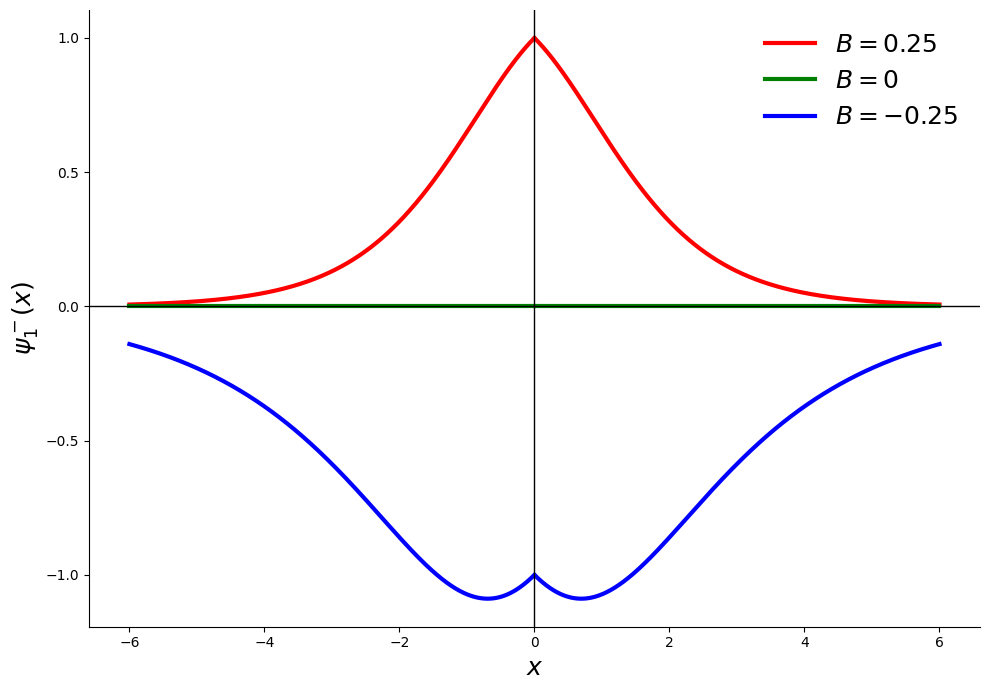}
        \caption{1st excited state for $\alpha=2B$, $A=3B$.}
    \label{F003a}
    \end{subfigure}
    \hfill 
    \begin{subfigure}[b]{0.45\textwidth}
        \centering
        \includegraphics[width=\textwidth]{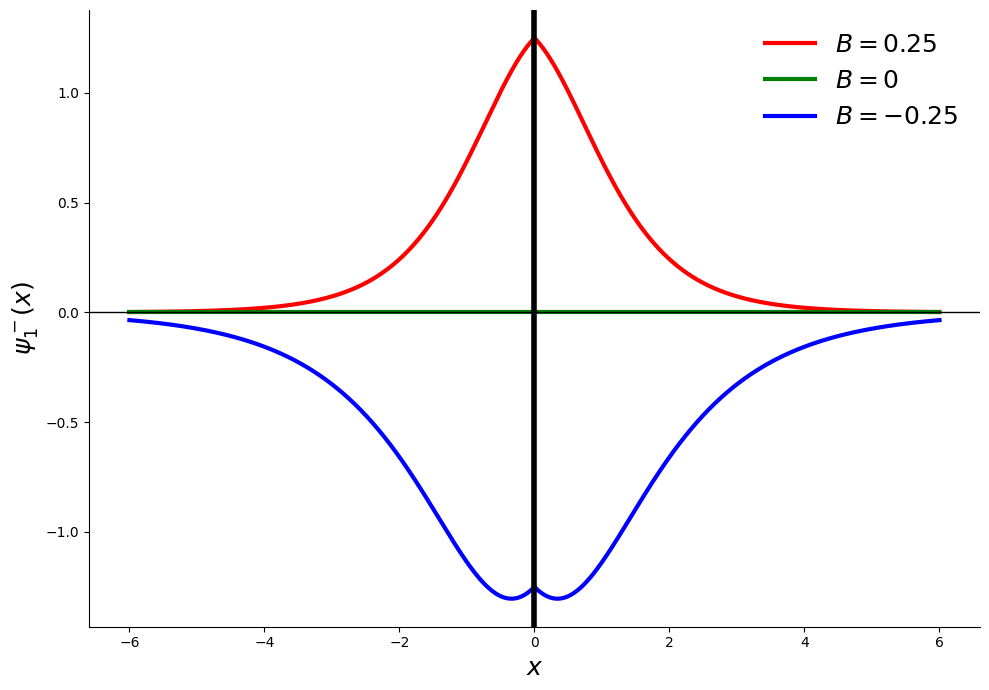}
        \caption{1st excited state for $\alpha=-3B$, $A=-4B$.}
        \label{F003b}
    \end{subfigure}
    \caption{The 1st excited state for the Rosen-Morse-type potential consistent with overall constraints. It vanishes for $B=0$, otherwise having a discontinuous slope determined by the sign of $B$.}
    \label{F003}
\end{figure}

Since the Rosen-Morse potential is exactly solvable \cite{RosenMorse1932}, the general expression of its eigenfunctions is well-known \cite{COOPER1995267}:
\begin{eqnarray}
    &&\psi_n^-(x)=N_n\left(1-\tanh(\alpha x)\right)^{s_+/2}\left(1+\tanh(\alpha x)\right)^{s_-/2}P_n^{(s_+,s_-)}\left(\tanh(\alpha x)\right);\nonumber\\
    &&s_\pm=\frac{A}{\alpha}-n\pm\frac{AB}{\alpha\left(A-n\alpha\right)},
\end{eqnarray}
with normalization constant $N_n$. $P_n^{(s_+,s_-)}\left(\tanh(\alpha x)\right)$ are the Jacobi polynomials determining the bound state energies that have already been obtained. However, for the present treatment, it is analytically economical to exploit the shape-invariance of the system \cite{gendenshtein1983,COOPER1995267}. Then the first excited state of $V_-(x)$ follows from Eq. \ref{Eq16a} with $a_1=\left(A_1,\,B_1\right)$.
Subsequently, the distribution of the wavefunction upon regularization takes the form,
\begin{equation}
    \psi_1^-(x)=
    \begin{cases}
        \varphi_I(x)=N_1^-\left[(A_1+A)\tanh(\alpha x)-(B_1+B)\right]\exp(B_1x)~\text{sech}^{A_1/\alpha}(\alpha x)  & \text{for}~ x<-\varepsilon, \\
        \varphi_{II}(x)=C\cos(px)+D\sin(px) & \text{for}~ -\varepsilon\leq x\leq\varepsilon, \\
       \varphi_{III}(x)=N_1^+\left[(A_1+A)\tanh(\alpha x)+(B_1+B)\right]\exp(-B_1x)~\text{sech}^{A_1/\alpha}(\alpha x)  & \text{for}~ \varepsilon<x.
    \end{cases}\label{E032}
\end{equation}
Similar to the ground state, implementing the boundary conditions at $x=\pm\varepsilon$ leads to two competing situations. Only the parity-even combination, shown in Fig. \ref{F003}, survives the DDF limit and abides by the parametric condition,
\begin{equation}
    \boxed{\frac{\Omega}{2}=B_1-\alpha\frac{A_1+A}{B_1+B}}.\label{E038A}
\end{equation}
The detailed calculations are shown in Appendix \ref{A1}. Coming from the boundary conditions in Eq. \ref{E033}, the above equation is indeed an additional constraint that is different from $\Omega=2B$. Since it derives directly from the superpotential, the latter is more fundamental. Therefore, the 1st excited state can exist only if both of these constraints agree, which requires that,
\begin{equation}
    A=\alpha\pm B\quad B=\Omega/2.
\end{equation}
This restriction is necessary for this state to satisfy the general discontinuity condition of Eq. \ref{Eq07} (Appendix \ref{A1}), as shape-invariance works only individually for the regular parts of the potential. 
This eigenfunction has been plotted in Fig. \ref{F003}, having parameter values subjected to the overall constraints, with the slope discontinuity at the origin controlled by the strength parameter $B$ of the DDF singularity as expected. 

\subsubsection{Symmetry restrictions and higher eigenstates}
The lack of parity symmetry of the two constituent systems of Eq, \ref{E003} (Fig. \ref{F004}) implies that the corresponding eigenstates do not possess definite parity. Consequently, the odd parity combinations of these eigenstates about $x=0$ are not continuous there. Therefore, only the even-parity combinations survive the boundary conditions, although the combined potential is parity-even. Subsequently, since $x=0$ is not an extremum of the constituent states in general, the combined state must have a dip or a peak there. This is qualitatively different from the widely observed dips when DDF was simply added to other parity-even potentials \cite{flugge2012practical,10.1119/1.9857,Belloni2014DeltaPotential}. In particular, this means a sharp peak at the origin given $B>0$\footnote{For $B<0$, this argument shifts to the superpartner $V_+(x)$.} for the ground state as it additionally satisfies,
\begin{equation}
    \left.\frac{d\psi_0^-}{dx}\right\vert_{x\to 0^\pm}=\mp B\varphi_0,\quad\varphi_0=\psi_0^-(0),
\end{equation}
following Eq. \ref{Eq08}. This further makes sense as a ground state cannot have any other local minimum. As for the excited states, since they are parity-even, 
\begin{equation}
    \left.\frac{d\psi_n^-}{dx}\right\vert^{x\to 0^+}_{x\to 0^-}\equiv 2\left.\frac{d\psi_n^-}{dx}\right\vert_{x\to 0^\pm}=\mp\Omega\varphi_n,\quad\varphi_n=\psi_n^-(0).
\end{equation}
Individually, $\Omega\neq 2B$ for $n> 0$ and either a dip or a sharp peak can appear at $x=0$. However, considering the ground state constraint $\Omega=2B$ simultaneously, these states eventually peak at the origin for $B>0$. Eventually, more and more constraints accumulate on the system parameters $(A,B,\alpha)$ as $n$ increases, limiting the number of eigenstates that can exist consistently. As a demonstration, we evaluate the next excited state below.

\paragraph*{Second excited state:}
Implementing the shape-invariant formula in Eq. \ref{Eq016b} for the second excited state,
with $a_2=\left(A_2,\,B_2\right)$ in this case, the distribution for the second excited state upon regularization takes the form,
\begin{eqnarray}
    &&\psi_2^-(x)=
    \begin{cases}
        \varphi_I(x)=N_2^-\left[C_1-C_2~\text{sech}^2(\alpha x)-C_3\tanh(\alpha x)\right]\exp(B_2x)~\text{sech}^{A_2/\alpha}(\alpha x)  & \text{for}~ x<-\varepsilon, \\
        \varphi_{II}(x)=C\cos(px)+D\sin(px) & \text{for}~ -\varepsilon\leq x\leq\varepsilon, \\
       \varphi_{III}(x)=N_2^+\left[C_1-C_2~\text{sech}^2(\alpha x)+C_3\tanh(\alpha x)\right]\exp(-B_2x)~\text{sech}^{A_2/\alpha}(\alpha x)  & \text{for}~ \varepsilon<x.
    \end{cases},\nonumber\\
    &&\text{where},\nonumber\\
    && C_1=(A_2+A_1)(A_2+A)+(B_2+B_1)(B_2+B),\nonumber\\
    && C_2=(A_2+A_1)(A_2+A+\alpha),\nonumber\\
    && C_3=(A_2+A_1)(B_2+B)+(A_2+A)(B_2+B_1).\label{E042}
\end{eqnarray}
The subsequent calculations are shown in Appendix \ref{A2}. The DDF limit again picks up the parity-even combination through the particular constraint condition,
\begin{equation}
   \boxed{\frac{\Omega}{2}=B_2-\alpha\frac{(A_2+A_1)(B_2+B)+(A_2+A)(B_2+B_1)}{(B_2+B_1)(B_2+B)-\alpha(A_2+A_1)}},\label{E048A}
\end{equation}
that conforms to the self-adjoint extension. When considered simultaneously with the previous constraints obtained for $\psi_{0,1}^-(x)$ we eventually get,
\begin{equation}
    \alpha=\begin{cases}
        0, 2B & \text{for}~A=\alpha+ B\\
        0,-3B & \text{for}~A=\alpha- B
    \end{cases}~,
\quad B=\frac{\Omega}{2}.\label{ConstFull}
\end{equation}
\begin{figure}[t]
    \centering
        \includegraphics[width=0.45\textwidth]{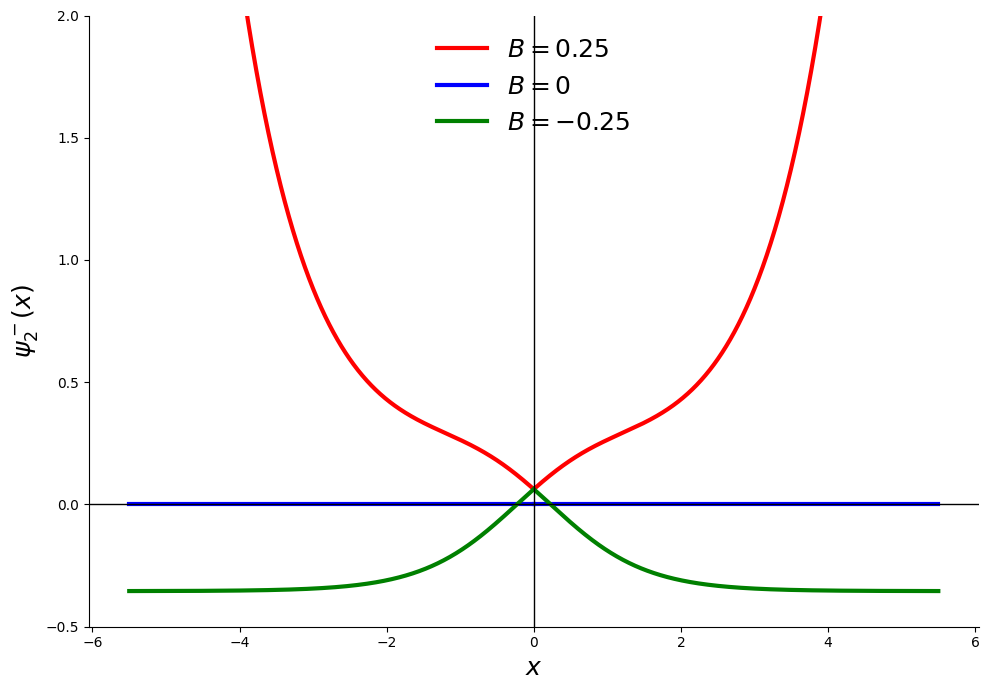}
    \caption{2nd excited state for the Rosen-Morse potential for $\alpha=-3B$ and $A=-4B$ only. The sign of $B$ still determines the slop at $x=0$ but non-trivial cases are not normalizable.}
    \label{F005}
\end{figure}
As a result, all the parameters of the system are fixed in terms of the strength of the DDF. Overlooking the trivial case of a constant potential for $\alpha=0$, which collapses all three eigenstates to the single exponentially decaying bound state \cite{griffiths2017introduction},
\begin{equation}
    \left.\psi_{0,1,2}^-(x;A,B,\alpha)\right\vert_{\alpha=0}\propto\exp\left(-B\vert x\vert\right),
\end{equation}
there are two notable possibilities: 
\begin{enumerate}
    \item For the combination $(B=\Omega/2,\,\alpha=2B,\,A=3B)$, the ground and the first excited have respective normalizations $N_0=\sqrt{\Omega}$ and $N_1=1/\sqrt{4\Omega}$, whereas the second excited state does not survive as the coefficients $C_{1,2,3}$ vanish identically. As for the respective spectra of the two disjoint parts $E_n=A^2+B^2-A_n^2-B_n^2$, both the lowest lying states are degenerate for the given parameter values with $E_{0,1}=0$. 
    They are plotted in Fig.s \ref{F002a} and \ref{F003a} respectively. 
    \item For the combination $(B=\Omega/2,\,\alpha=-3B,\,A=-4B)$, the situation remains similar. The lowest two states are again degenerate with $E_{0,1}=0$, having normalization constants $N_0=\sqrt{5\Omega}/2$ and $N_1=1/\sqrt{5\Omega}$. These states are shown in Fig.s \ref{F002b} and \ref{F003b} respectively. The second excited state, although corresponds to non-trivial values $C_1=-\Omega^2$ and $C_{2,3}=-(5/4)\Omega^2$, turns out to be non-normalizable since $\Omega>0$. This situation is depicted in Fig. \ref{F005}, with even individual parts corresponding to each segment of the potential are themselves being non-normalizable. 
\end{enumerate}
We can safely infer that there will not be any more excited states since all the parameters have been exhausted by the constraints. From the shape-invariance point-of-view, the absence of a feasible 2nd excited state further supports this assertion. This is unlike the bounding systems considered before in addition to a DDF singularity \cite{Belloni2014DeltaPotential} that supported the full spectra. The situation will be even more restrictive for the superpartner $V_+(x)$. Notice that the eigenfunctions actually vanish for $B=0$ instead of becoming smooth, as seen in Fig.s \ref{Fig002}, \ref{F003} and \ref{F005}, when all the constraints are accounted for. This is distinctly different from when the DDF is outside the isospectral structure \cite{flugge2012practical,Belloni2014DeltaPotential} with just the dip in the even states being removed. The built-in discontinuity (singularity) of the isospectral structure is responsible for this unique property that prohibits analytical continuation to the case of a smooth potential. This is reflected in the constraint condition of Eq. \ref{E038A} which is singular for $B=0$. Even the regularization mechanism itself breaks down since $p^2\varepsilon\to 0$ in that case.

\subsection{The superpartner $V_+(x)$}
For the sake of completion, demonstrating the highly constrained nature of this system, we now look at the isospectral partner potential $V_+$ that has the form,
\begin{equation}
    V_+(x\gtrless 0)=A^2+B^2-A\left(A-\alpha\right)\text{sech}^2(\alpha x)\pm 2AB\tanh(\alpha x).\label{E050}
\end{equation}
The $2B\delta(x)$ contribution can now be modeled as a {\it barrier} of height $V_0$ and width $2\varepsilon$ so that $V_0\varepsilon\to B$ in the DDF limit. Therefore, we have the full potential profile as,
\begin{equation}
    V_-(x)=\begin{cases}
A^2+B^2-A\left(A-\alpha\right)\text{sech}^2(\alpha x)- 2AB\tanh(\alpha x) & \text{for}~ x<-\varepsilon, \\
V_0 & \text{for}~ -\varepsilon\leq x\leq\varepsilon, \\
A^2+B^2-A\left(A-\alpha\right)\text{sech}^2(\alpha x)+ 2AB\tanh(\alpha x) & \text{for}~ \varepsilon<x.
\end{cases}\label{E051}
\end{equation}

\paragraph{Ground state:} The ground states on both sides will emerge from the usual isospectral mapping as,
\begin{equation}
    \psi_0^+(x\gtrless 0;a_0)=\left(\frac{d}{dx}+W(x\gtrless 0;a_0)\right)\psi_1^-(x\gtrless 0;a_0).\label{E053}
\end{equation}
Then, taking account of Eq. \ref{E032}, we get the ground state distribution of $V_+$ as,
\begin{eqnarray}
    &&\psi_0^+(x)=
    \begin{cases}
        \varphi_I(x)=M_0^-\left[-D_1+D_2~\text{sech}^2(\alpha x)+D_3\tanh(\alpha x)\right]\exp(B_1x)~\text{sech}^{A_1/\alpha}(\alpha x)  & \text{for}~ x<-\varepsilon, \\
        \varphi_{II}(x)=C\exp(px)+D\exp(-px) & \text{for}~ -\varepsilon\leq x\leq\varepsilon, \\
       \varphi_{III}(x)=M_0^+\left[-D_1+D_2~\text{sech}^2(\alpha x)-D_3\tanh(\alpha x)\right]\exp(-B_1x)~\text{sech}^{A_1/\alpha}(\alpha x)  & \text{for}~ \varepsilon<x.
    \end{cases},\nonumber\\
    &&\text{where},\nonumber\\
    && D_1=A_1^2+B_1^2-A^2-B^2,\nonumber\\
    && D_2=\left(A_1^2-A^2\right)+\alpha(A_1+A)\nonumber\\
    && D_3=(A_1+A)(B_1-B)+(A_1-A)(B_1+B)\equiv 0.\label{E054}
\end{eqnarray}
Here, $p=\sqrt{V_0-E}\in\mathbb{R}$ serves as the momentum of a locally decaying state within the barrier. 

\paragraph*{} The subsequent calculations for the boundary conditions are carried out in Appendix \ref{B1}. After resolving the two possible cases, which qualitatively differ from those for the eigenstates of $V_-(x)$, the DDF limit again singles out the even-parity combination corresponding to a constraint $\boxed{\Omega/2=B_1}$. Compatibility with the constraints from $V_-$ immediately restricts the existence of $\psi_0^+(x)$. In particular, implementing $\Omega/2=B$ for $\psi_0^-(x)$ simultaneously, one uniquely gets the trivial result $\alpha=0$. Consequently, subject to the restricted parameter values of Eq. \ref{ConstFull} obtained for $V_-(x)$, no nontrivial results can be retained. 

\paragraph{1st excited state:} For a consistency check, we consider the 1st excited state for $V_+(x)$ obtained using the shape-invariant formula: 
\begin{equation}
    \psi_1^+(x\gtrless 0;a_0)=\left(\frac{d}{dx}+W(x\gtrless 0;a_0)\right)\psi_2^-(x\gtrless 0;a_1).
\end{equation}
Upon evaluation, this state is distributed across the regularized barrier as,
\begin{eqnarray}
    &&\psi_1^+(x)=
    \begin{cases}
        \varphi_I(x)=M_1^-\left[E_1-E_2~\text{sech}^2(\alpha x)-E_3\tanh(\alpha x)\right]\exp(B_2x)~\text{sech}^{A_2/\alpha}(\alpha x)  & \text{for}~ x<-\varepsilon, \\
        \varphi_{II}(x)=C\exp(px)+D\exp(-px) & \text{for}~ -\varepsilon\leq x\leq\varepsilon, \\
       \varphi_{III}(x)=M_1^+\left[-E_1+E_2~\text{sech}^2(\alpha x)-E_3\tanh(\alpha x)\right]\exp(-B_2x)~\text{sech}^{A_2/\alpha}(\alpha x)  & \text{for}~ \varepsilon<x.
    \end{cases},\nonumber\\
    &&\text{where},\nonumber\\
    && E_1=C_1(B_2-B)+C_3(A_2-A),\nonumber\\
    && E_2=C_2(B_2-B)+C_3(A_2-A+\alpha)\nonumber\\
    && E_3=C_1(A_2-A)+C_3(B_2-B).\label{E061}
\end{eqnarray}
with $C_{1,2,3}$ coming out of Eq. \ref{E042}. Subsequently implementing the boundary conditions in Appendix \ref{B2} eventually picks out a parity-even combination with a complicated-looking constraint condition,
\begin{equation}
    \boxed{\frac{\Omega}{2}=\alpha\frac{(A_2+A_1)\left(A_2^2+B_2^2-A^2-B^2\right)}{(B_2+B_1)\left(B_2^2-B^2\right)-4\alpha(AB+A_1B_2)}-B_2}.\label{E066A}
\end{equation}
This constraint eventually eludes the trivial case $\alpha=0$ in light of the prior ones from $V_-(x)$, and the present eigenfunction identically vanishes. This behavior is expected to continue further, and we infer that the superpartner $V_+(x)$ does not survive the boundary conditions, which in turn supplements the degeneracy in the surviving states of $V_-(x)$ with vanishing energy.

\paragraph*{} Survival of the entire spectra subject to selective distortions when DDF was not integrated isospectrally
\cite{Belloni2014DeltaPotential,10.1119/1.9857,flugge2012practical} can be attributed to the two boundary conditions at $x=\pm\varepsilon$ involving the {\it same} pair of systems. Since the two components of the present potential are different, each eigenstate requires additional compatibility under the DDF limit that successively constrains the system parameters. It can be inferred that the number of surviving states can increase with the number of system parameters as more constraints can be accommodated, serving as a {\it spectral regulator}. Additionally, unlike the ones for the previously studied cases where the DDF was included perturbatively \cite{Belloni2014DeltaPotential,10.1119/1.9857}, the states of these isospectrally singular systems are exact since the discontinuous superpotential was maintained throughout.


\section{Presence of additional inverse-square interaction}\label{S5}
Unlike the DDF potential, which has a self-adjoint extension implementing a boundary condition at its location, a smoothly varying yet singular potential would enforce a power-law behavior of the wavefunction near the singularity \cite{landau1977quantum}. Although the particular self-adjoint extension for the latter system depends on the exact form of the singularity, the derivative of the wavefunction is expected to be singular since the wavefunctions themselves are \cite{RevModPhys.43.36}. In particular, the $\lambda/x^2$-type potential shows this behavior but in a certain parameter range ($\lambda<3/4$) \cite{reed1975methods}. The exact behavior of the corresponding wavefunction is governed by its self-adjoint extensions that effectively manifest through the strength $\lambda$. The system only admits a continuum of hard scattering states for $\lambda>3/4$ \cite{Gitman_2010} whereas a $U(2)$ parameter family of extensions decides either a scattering continuum or a single bound state for $-1/4\leq\lambda<3/4$ \cite{PhysRevA.89.022113}. The system supports an infinite, unbounded tower of bound states for $\lambda<-1/4$ \cite{PhysRevA.67.042712}. 

\paragraph*{} Combining such a system with a DDF potential can seldom lead to nontrivial spectra. The simplest construction can be with the superpotential,
\begin{equation}
    W_{ISD}(x)=\frac{C}{x}+B~\text{Sgn}(x),\label{EqISD}
\end{equation}
which is itself singular. Apart from the DDF singularity $-2B\delta(x)$, the corresponding potential segments,
\begin{equation}
    V_-(x\gtrless 0)=\frac{C(C+1)}{x^2}+\frac{2BC}{\vert x\vert}+B^2,
\end{equation}
individually supports shape-invariance with the parameterization and subsequent spectra,
\begin{eqnarray}
    &&C_n=C-n,\quad B_n=\frac{BC}{C_n},\quad R(a_n)=B^2C^2\left(\frac{1}{C_n^2}-\frac{1}{C_{n+1}^2}\right),\nonumber\\
    && E_n=B^2\left[1-\left(\frac{C}{C-n}\right)^2\right], \label{EqNew01}
\end{eqnarray}
respectively, with the latter being bounded. The strength of the $x^{-2}$ singularity is now identified as $\lambda=C(C+1)$\footnote{For real $C$, this isospectral parameterization prohibits access to the 'fall-of-the-center' domain $\lambda<-1/4$.}, and thus the hard-scattering continuum condition now translates to $C>1/2$. This is reflected in the individual ground states generated from $W_{ISD}(x)$,
\begin{equation}
    \psi_0^-(x\gtrless 0)\sim\vert x\vert^{-C}\exp(\mp Bx),\label{GSISD} 
\end{equation}
which are localized and can be normalizable only if $C<1/2$ and $B>0$ if $x=0$ is included. The nontrivial spectra in Eq. \ref{EqNew01} highlight the effect of a constant shift $\pm B$ in the superpotential of the $x^{-2}$-potential, the latter having only a single bound state for $C<1/2$ \cite{PhysRevA.89.022113} with $C_n=C-n$.

\paragraph*{}  However, the two competing singularities in the potentials require distinct regularization schemes, which is evident as the erstwhile $\varepsilon$-expansion in the case of the DDF is singular with inverse powers of $x$ in the potential. Further, since $W_{ISD}(x)$ itself will be singular, the excited states obtained through shape-invariance will be increasingly so. A suitably regularized case could be,
\begin{equation}
    W_{ISDR}(x)=\frac{C}{x-\alpha~\text{Sgn}(x)}+B~\text{Sgn}(x),\label{EqNew02}
\end{equation}
with a cut-off regularization to the $x^{-2}$-term, generating a symmetric potential about the DDF singularity\footnote{As we have seen, reflection symmetry about the DDF is needed to satisfy the derivative discontinuity condition.}. However, such a parameterization ruins any possibility of shape-invariance due to the regularizing $x\pm\alpha$ term in the denominator.  

\begin{figure}[t]
\centering
  \includegraphics[width=0.5\linewidth]{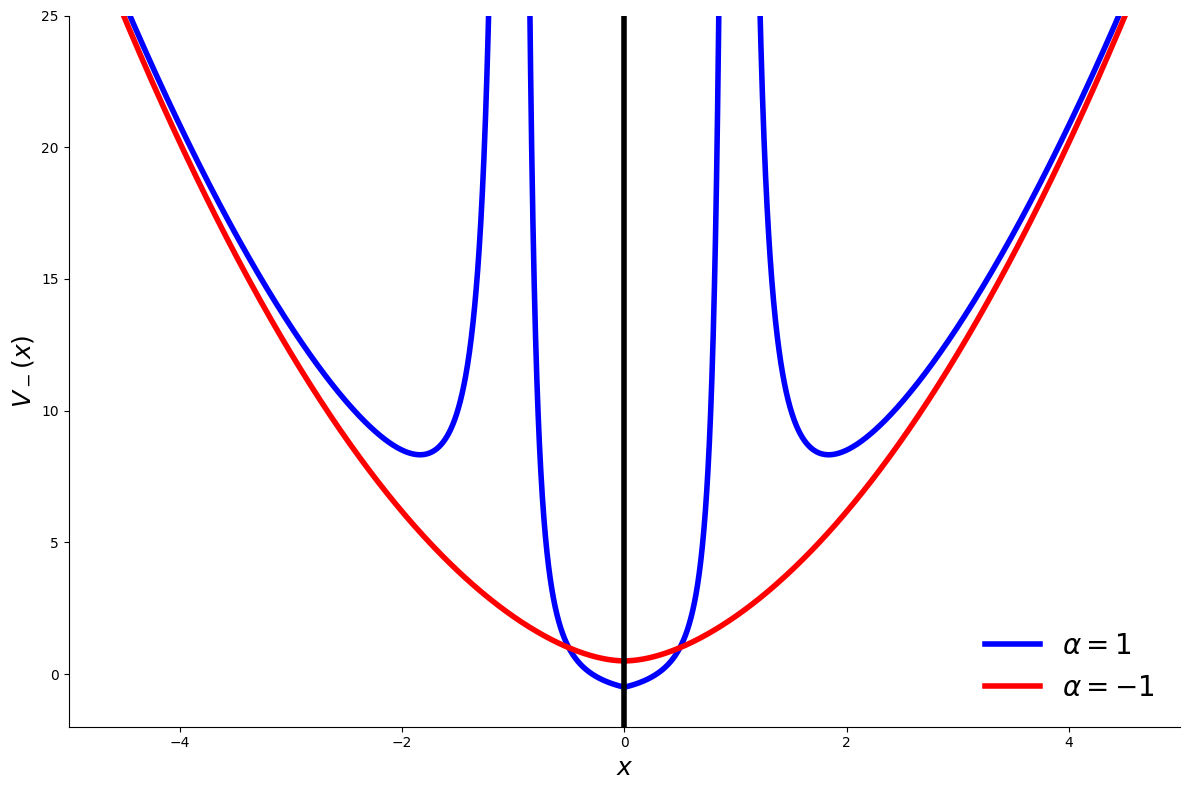}
  \caption{$V_-(x)$ for $A=1$, $C=0.5=B$ with presence ($\alpha=1$, blue) and absence ($\alpha=-1$, red) of the $\sim x^{-2}$-singularity.}
\label{F010}
\end{figure}

\paragraph*{} We, however, directly consider the case when an additional Harmonic trap is present. This is motivated by the well-known and exactly solvable Calogero-type \cite{Calogero1971} model,
\begin{equation}
    W_C(x)=Ax+\frac{C}{x},
\end{equation}
that supports an oscillator-type equispaced spectrum of $E_n=4An$ with shape-invariance parameterized as $\left(A_n=A,\,C_n=C-n\right)$. The corresponding bound states are \cite{10.1063/1.1664820},
\begin{equation}
    \psi_n^-\sim\vert x\vert^{-C}\exp\left(-Ax^2/2\right)L_n^{-B-1/2}\left(Ax^2\right),\label{EqLagurre}
\end{equation}
with associated Laguerre polynomials \cite{GradshteynRyzhik2007} $L_n^a(x)$, which are valid for $C<1/2$ \cite{GoldmanKrivchenkov2012} with genuine and broken SUSY sectors \cite{JEVICKI198455,PANIGRAHI1993251}. The isospectral inclusion of the DDF to this system,
\begin{equation}
    W_{CD}=Ax+\frac{C}{x}+B~\text{Sgn}(x).
\end{equation}
is expected to control the nontrivial spectrum, as in the previous cases. However, the DDF term itself maims the n\"aive shape-invariance since the spectrum collapses as $R(a_n)=0$ for the only possible parameterization of $\left(A_n=(-1)^nA,\,C_n=(-1)^nC,\,B_n=(-1)^nB\right)$, unlike when the harmonic trap was absent (Eq. \ref{EqNew01})\footnote{This is due to a larger number of parametric constraints implemented by shape-invariance}. As stated before, a cut-off regularization on the $x^{-2}$ term of the form,
\begin{eqnarray}
    &&W_{\text{CSD}}(x)=Ax+\frac{C}{x-\alpha~\text{Sgn}(x)}+B~\text{Sgn}(x),\nonumber\\
    &&V_\pm(x)=\left(Ax+\frac{C}{x-\alpha~\text{Sgn}(x)}+B~\text{Sgn}(x)\right)^2\pm\left[A-\frac{C}{\left(x-\alpha~\text{Sgn}(x)\right)^2}+2\left(-\frac{C}{\alpha}+B\right)\delta(x)\right].\label{EIS01}
\end{eqnarray}
yields a parity-even segmented potential, as shown in Fig. \ref{F010}, but ruins shape-invariance. Interestingly, the $x^{-2}$  singularity does not appear for $\alpha<0$ due to the segmented nature of the potential, escaping the need for subsequent self-adjoint extension and leading to a wider acceptable parametric range for continuous eigenfunctions. 

\paragraph*{} Multiple bound states are feasible here due to the harmonic confinement, although they will not follow the form of Eq. \ref{EqLagurre} due to the additional $\sim\left(\vert x\vert-\alpha\right)^{-1}$ term in the potential. The ground state can be readily obtained from the superpotential $W_{\text{CSD}}(x)$ as,
\begin{equation}
    \psi_0^-(x\gtrless 0)=N_0\vert x\mp\alpha\vert^{-C}\exp\left(-\frac{y_\pm^2}{2A}\right),\quad y_\pm=Ax\pm B,
\end{equation}
which is symmetric about the DDF. Then the usual limiting procedure around $x=0$ leads to the constraint,
\begin{equation}
    \boxed{\frac{\Omega}{2}=-\frac{C}{\alpha}+B}.\label{EqBC01}
\end{equation}
The relevant details are given in Appendix \ref{D}. Clearly, the two singularities do not remain mutually inert since the isospectral structure mixes the respective parameters through the $2BC/\left(\vert x\vert-\alpha\right)$ term in the potential. Moreover, they cannot coincide sensibly as the cut-off regularization cannot be removed ($\alpha\to 0$) in a consistent way. Therefore, the only feasible way out is to consider $\alpha$ as a physical parameter so that the two singularities are physically separated. 

\begin{figure}[t]
\begin{subfigure}{.475\linewidth}
  \includegraphics[width=\linewidth]{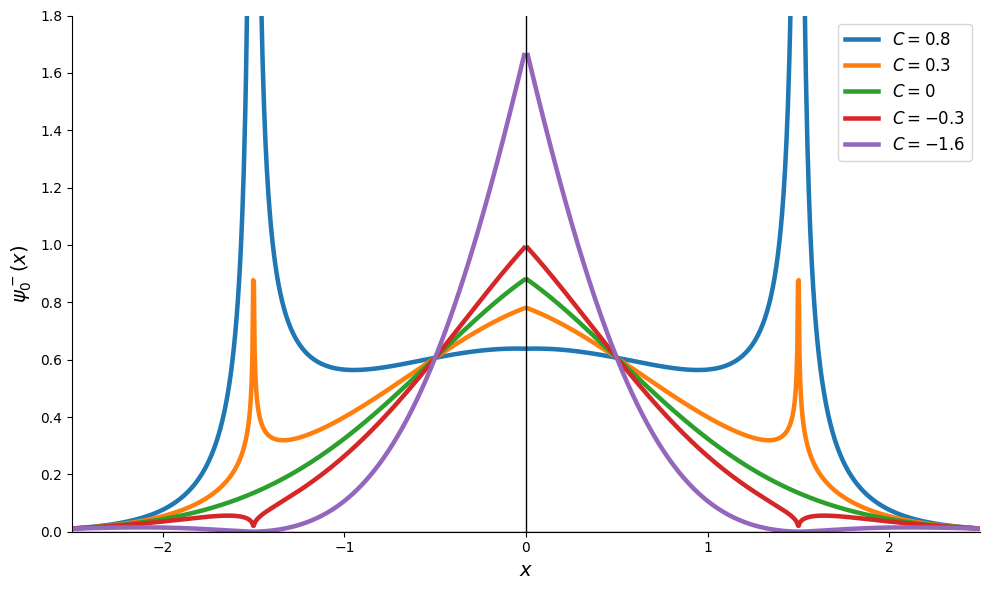}
  \caption{Ground state for $A=1$, $\alpha=1.5$ and $B=0.5$.}
  \label{F011a}
\end{subfigure}\hfill 
\begin{subfigure}{.475\linewidth}
  \includegraphics[width=\linewidth]{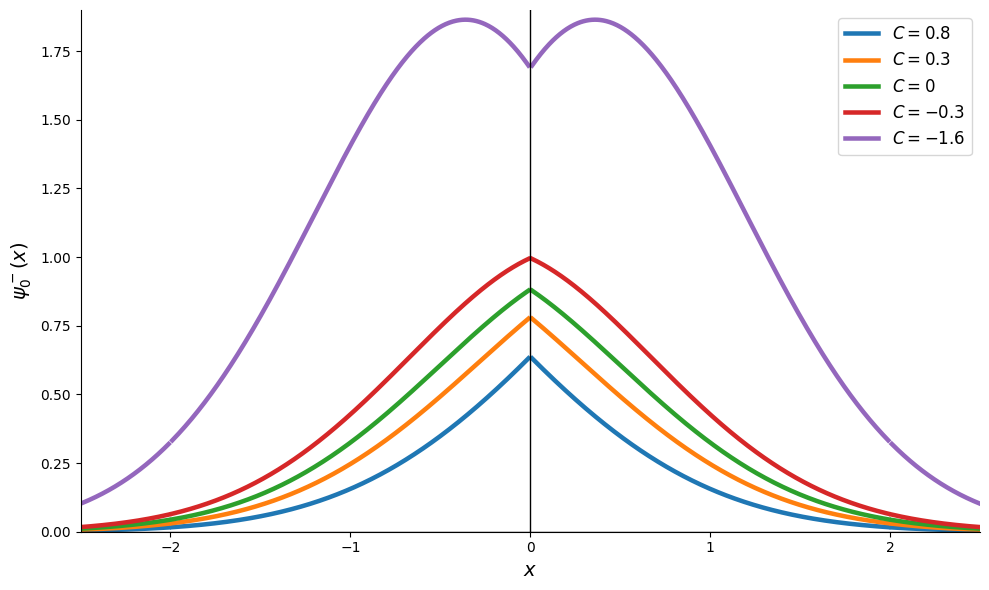}
  \caption{Ground state for $A=1$, $\alpha=-1.5$ and $B=0.5$.}
  \label{F011b}
\end{subfigure}
\caption{ Ground state with different analytic structures based on the $x^{-2}$-strength $C$ for the parametric phases $\alpha\gtrless 0$.}
\label{F011}
\end{figure}

\paragraph*{} The ground state is plotted in Fig. \ref{F011} for different `strengths' $C$ of the inverse-power singularities. When they manifest ($\alpha>0$), the respective parameter phases \cite{GoldmanKrivchenkov2012,Gitman_2010,PhysRevA.89.022113,PhysRevA.67.042712} pertaining to the non-trivial self-adjointness also manifest (Fig. \ref{F011a}). The ground state is not square-integrable for $C> 1/2$, hinting at hard scattering at $x=\pm\alpha$. But it is singular yet normalizable For $1/2>C>0$ and regains the non-singular behavior of Section \ref{S3} at $C=0$. For $C<0$, the normalization wavefunction shows finite discontinuities at $x=\pm\alpha$, the latter disappearing beyond $C=-1/2$, thereby returning to the hard scattering limit. Throughout, the derivative discontinuity at $x=0$ is evident. In the absence of the inverse-power singularities ($\alpha<0$), Fig. \ref{F011b} shows smooth behavior of the ground state, except for the usual one at $x=0$ owing to the DDF for all values of $C$, which is similar to that of Section \ref{S3}.

\paragraph*{} Since the singular term in $W_{\text{CSD}}(x)$ prohibits shape-invariance for $\alpha\neq 0$, which being necessary for a sensible outcome (Eq. \ref{EqBC01}), the excited states must be obtained by directly solving the corresponding Schr\"odinger equation, which is beyond the present scope. However, it can still be inferred that the excited states will produce additional constraints, eventually terminating the spectrum at some level. The absence of symmetry about $x=0$ of the individual segments of the potential in Eq. \ref{EIS02} supports the same conclusion, as we have observed previously. It is safe to say that the isospectral inclusion of DDF in a system with other singularities results in a constrained spectrum, which is further restricted by competing singularity effects. such a situation follows from the corresponding self-adjoint extensions, which are mutually independent.

\section{Discussions and conclusions}\label{S6}
We analyzed the role of the DDF singularity when it is embedded in the isospectral structure through a discontinuous superpotential, unlike previous studies. The system naturally splits into two distinct isospectral potentials across the location of the DDF; connected through boundary conditions that manifest the self-adjoint extension due to the DDF. The effect of such a singularity on the bound spectra is studied when the overall composite system is parity-even, even if the constituent systems on either side of the DDF are not. Since the superpotential for such a system is necessarily discontinuous, the algebraic evaluation of the eigenfunctions through shape-invariance necessarily requires a finite well/barrier-type regularization of the DDF. As a result, each composite eigenfunction is subjected to two usual but independent boundary conditions, imposing an additional parametric constraint. Each such constraint demands compatibility among the two segments of the particular eigenfunction that emerge from two different isospectral structures. Therefore, these independent constraints accumulate with every excited state and restrict the spectra depending on the number of free parameters in the system. In particular, the isospectral merger of the DDF with a harmonic trap turns out to be a single-level system, whereas a Morse-type potential with the DDF supports only two degenerate states of zero energy. 

\paragraph*{}Since the $\text{Sgn}(x)$ function is mandatory in the superpotential, the individual potentials on either side of the DDF can never be symmetric about its location. This rules out any overall eigenfunction that is odd about that location since it cannot be continuous there. Subsequently, the surviving overall eigenfunctions are always even about the DDF location, with a dip/peak there owing to the derivative discontinuity. These characteristics are evident in all the systems considered here, which is unlike the usual parity-even potentials with alternating even/odd eigenstates; the additional presence of the DDF in the latter only affects the even ones. In the oscillator potential with DDF, the two halves themselves are parity-even but about different points, whereas the two segments of the Rosen-Morse structure with the DDF do not possess any definite parity. However, undermining the compatibility between the individual constraints, all the eigenstates evaluated for both systems are parity-even, with their slope being discontinuous at $x=0$. Thus, the DDF `glues together' a pair of twin isospectral systems in a highly restrictive way, having a topologically nontrivial quantum interface. Interestingly, the lack of symmetry in the two parts of the Rosen-Morse structure led to two allowed states when the constraints were compared, whereas the individual symmetries of the two segments of the oscillator system allowed only one. The cumulative constraints further obstruct the supersymmetric structure itself, as no eigenstate is allowed for the respective superpartners containing an isospectral DDF scatterer.  

\paragraph*{}Finally, the isospectral coexistence of DDF with any other singularity is found to be highly incompatible, essentially due to their conflicting individual regularization. Considering the case of the $x^{-2}$-type singularity in detail, it is seen that the two singularities cannot coincide. Separating them in a way such that the overall potential is symmetric about the DDF, with an additional harmonic trap forming a Calogero-type set-up for a bound spectrum, shape-invariance is necessarily lost. As a result, obtaining the excited states is challenging, given that isospectrality necessarily introduces an additional $x^{-1}$-singularity. Even the ground state does not survive the coincidence limit of the two singularities, although one can choose to avoid the inverse-power singularities through a suitable parameter choice. This suggests a general incompatibility between distributional and power-law singularities.

\paragraph*{} An isospectral potential containing the DDF can underline a mechanism to engineer systems with finite spectra that could be controlled by system parameters, particularly in the case of localized defects when coupled structurally. More formally, similar models can provide interplay between self-adjoint extensions and supersymmetric structures in singular quantum systems. The framework developed here opens avenues for controlled spectral design and provides deeper insight into the role of singular interactions in quantum mechanics. We plan to study the algebraic structure of such systems in the near future.

\appendix
\section{Reducing $D_n(x)$s to usual oscillator eigenstates}\label{A00}
Subjecting the general eigenfunction of Eq.s \ref{ES05a} to the boundary conditions at $x=\pm\varepsilon$ yields the following expressions,
\begin{eqnarray}
    &&pC=\begin{cases}
        N_{n_-}^-\left[p\cos(p\varepsilon)D_{n_-}(-\phi)+\sin(p\varepsilon)\left\{\sqrt{2A}n_-D_{n_--1}(-\phi)+\theta D_{n_-}(-\phi)\right\}\right], & \text{at}~x=-\varepsilon\\
        N_{n_+}^+\left[p\cos(p\varepsilon)D_{n_+}(\phi)-\sin(p\varepsilon)\left\{\sqrt{2A}n_+D_{n_+-1}(\phi)-\theta D_{n_+}(\phi)\right\}\right], & \text{at}~x=\varepsilon
    \end{cases},\label{EqA01a}\\
    &&pD=\begin{cases}
        N_{n_-}^-\left[-p\sin(p\varepsilon)D_{n_-}(-\phi)+\cos(p\varepsilon)\left\{\sqrt{2A}n_-D_{n_--1}(-\phi)+\theta D_{n_-}(-\phi)\right\}\right], & \text{at}~x=-\varepsilon\\
        N_{n_+}^+\left[p\sin(p\varepsilon)D_{n_+}(\phi)+\cos(p\varepsilon)\left\{\sqrt{2A}n_+D_{n_+-1}(\phi)-\theta D_{n_+}(\phi)\right\}\right], & \text{at}~x=\varepsilon
    \end{cases},\label{EqA01b}\\
    &&\text{where}\quad\theta=A\varepsilon+B\quad\text{and}\quad\phi=\sqrt{\frac{2}{a}}\theta.\nonumber
\end{eqnarray}
In doing so, the recursion relation,
\begin{equation}
    \frac{d}{dz}D_n(z)=nD_{n-1}(z)-\frac{z}{2}D_n(z),
\end{equation}
for the parabolic cylinder functions is utilized. To be consistent, both the expressions for $C$ and those for $D$ must be equal. Since the parabolic cylinder function further satisfy,
\begin{equation}
    D_n(-z)=(-1)^nD_n(z)+i\frac{\sqrt{2\pi}}{\Gamma(-n)}\exp\left(in\frac{\pi}{2}\right)D_{-n-1}(iz),\label{EqA02}
\end{equation}
wherein $D_n(z)$ and $D_{-n-1}(iz)$ are independent functions, this equality among the two expressions of $C$ implies,
\begin{eqnarray}
    &&N_{n_+}^+D_{n_+}(\phi)=(-1)^{n_-}N_{n_-}^-D_{n_-}(\phi)\nonumber\\
    &&ip\cos(p\varepsilon)D_{-n_--1}(i\phi)+\sin(p\varepsilon)\left\{-\sqrt{2A}D_{-n_-}(i\phi)+i\theta D_{-n_--1}(i\phi)\right\}=0.\label{EqA03a}
\end{eqnarray}
On the other hand, equating two expressions of $D$ leads to,
\begin{eqnarray}
    &&N_{n_+}^+D_{n_+}(\phi)=(-1)^{n_-+1}N_{n_-}^-D_{n_-}(\phi)\nonumber\\
    &&ip\sin(p\varepsilon)D_{-n_--1}(i\phi)+\cos(p\varepsilon)\left\{-\sqrt{2A}D_{-n_-}(i\phi)+i\theta D_{-n_--1}(i\phi)\right\}=0.\label{EqA03b}
\end{eqnarray}
Just like the previous systems in Section \ref{S3}, since these two sets of conditions are mutually incompatible, we conclude that either $D=0$ or $C=0$. In particular, since the parabolic cylinder functions with different indices are linearly independent, the only way the firsts of Eq.s \ref{EqA03a} and \ref{EqA03b} can be valid is $n_+=n_-\equiv n$. Further, considering the relation,
\begin{equation}
    D_n(iz)=\frac{\Gamma(1+n)}{\sqrt{2\pi}}\left[e^{in\pi/2}D_{-n-1}(z)+e^{-in\pi/2}D_{-n-1}(-z)\right],\label{EqA04}
\end{equation}
with $D_n(\pm z)$s being mutually independent, substitution in the second of Eq. \ref{EqA03a} leads to,
\begin{equation}
    p\cos(p\varepsilon)D_{n_+}(\phi)=\sin(p\varepsilon)\left\{\sqrt{2A}n_+D_{n_+-1}(\phi)-\theta D_{n_+}(\phi)\right\},
\end{equation}
where the first of Eq.s \ref{EqA03a} is used. As a result, $C$ vanishes identically. Similarly, substituting Eq. \ref{EqA04} in the second of Eq.s \ref{EqA03b} and using the first, we get $D=0$. Thus, for non-trivial $C$ and $D$, the second part of $D_n(-z)$ in Eq. \ref{EqA02} should not exist. This is possible only if, 
\begin{equation}
    n_+=n_-\equiv n\in\mathbb{Z}_{\geq 0},
\end{equation}
reducing the parabolic cylinder functions to the form in Eq. \ref{HerPol} with Hermite polynomials. This result also naturally arises from the boundary conditions at $x=\pm\varepsilon$ that imbibes the parity-even nature of the system.

\section{Eigenstates for the singular Rosen-Morse potential $V_-(x)$}\label{A}
\subsection{Ground state}\label{A0}
On applying the boundary conditions at $x=\pm\varepsilon$ to the wavefunctions of Eq. \ref{E019}, the coefficients of the standing wave can be expressed as:
\begin{eqnarray}
    &&pC=\begin{cases}
        \frac{\varphi_I(-\varepsilon)}{p}\left[p\cos(p\varepsilon)+\sin(p\varepsilon)\left\{B+A\tanh(\alpha\varepsilon)\right\}\right], & x=-\varepsilon\\
        \frac{\varphi_{III}(\varepsilon)}{p}\left[p\cos(p\varepsilon)+\sin(p\varepsilon)\left\{B+A\tanh(\alpha\varepsilon)\right\}\right], & x=\varepsilon
    \end{cases},\nonumber\\
    &&pD=\begin{cases}
        -\frac{\varphi_I(-\varepsilon)}{p}\left[p\sin(p\varepsilon)-\cos(p\varepsilon)\left\{B+A\tanh(\alpha\varepsilon)\right\}\right], & x=-\varepsilon\\
    \frac{\varphi_{III}(\varepsilon)}{p}\left[p\sin(p\varepsilon)-\cos(p\varepsilon)\left\{B+A\tanh(\alpha\varepsilon)\right\}\right], & x=\varepsilon
    \end{cases},
\end{eqnarray}
where $\varphi_{I,III}$ is given in Eq. \ref{E019}. As before, on comparing their expressions at $x=\pm\varepsilon$, $C$ and $D$ cannot be simultaneously non-zero, leading to two possibilities:
\begin{itemize}
    \item[{\bf Case 1}:] $N_0^-=N_0^+=N_0$ with $D=0$. The latter implies,
    \begin{equation}
        p\tan(p\varepsilon)=B+A\tanh(\alpha\varepsilon),\label{E025}
    \end{equation}
    leading to,
    \begin{equation}
        C=N_0\exp(-B\varepsilon)~\text{sech}^{A/\alpha}(\alpha\varepsilon)\sec(p\varepsilon).\label{E026}
    \end{equation}
    Thus $C$ has the same sign as $\varphi_{I,III}$ at the boundaries, leading to a {\bf parity-even} eigenstate.
    \item[{\bf Case 2}:] $N_0^-=-N_0^+=N_0$ with $C=0$. The latter implies,
    \begin{equation}
        -p\cot(p\varepsilon)=B+A\tanh(\alpha\varepsilon),\label{E027}
    \end{equation}
    leading to,
    \begin{equation}
        D=-N_0\exp(-B\varepsilon)~\text{sech}^{A/\alpha}(\alpha\varepsilon)~\text{cosec}(p\varepsilon).\label{E028}
    \end{equation}
    Thus $D$ has the same sign as $\varphi_I$ at the boundaries, but $\varphi_{III}$ has the opposite signs, leading to a {\bf parity-odd} eigenstate.
\end{itemize}

\paragraph*{} The DDF limit $\varepsilon\to 0,~p\to\infty;~p^2\varepsilon\to\Omega/2$
for Case 1 leads to $C=N_0$ along-with the constraint condition: $\boxed{\Omega=2B}$, representing the singularity $-2B\delta(x)$ at the origin. As for {\bf Case 2}, $B$ is constrained to vanish in the DDF limit. This parity-odd wavefunction does not even meet at the origin and can be rejected. 

\subsection{1st excited state}\label{A1}
Similarly to the ground state, the boundary conditions on the wavefunction in Eq. \ref{E032} lead to,
\begin{eqnarray}
    &&C=
    \begin{cases}
        \frac{N^-_1}{p}\exp(-B_1\varepsilon)~\text{sech}^{A_1/\alpha}(\alpha\varepsilon)\cos(p\varepsilon)\left[Pp+Q\tan(p\varepsilon)\right]  & \text{at}~ x=-\varepsilon, \\
        -\frac{N^+_1}{p}\exp(-B_1\varepsilon)~\text{sech}^{A_1/\alpha}(\alpha\varepsilon)\cos(p\varepsilon)\left[Pp+Q\tan(p\varepsilon)\right]  & \text{at}~ x=\varepsilon,
    \end{cases}\nonumber\\
    &&D=\begin{cases}
        -\frac{N_1^-}{p}\exp(-B_1\varepsilon)~\text{sech}^{A_1/\alpha}(\alpha\varepsilon)\sin(p\varepsilon)\left[Pp-Q\cot(p\varepsilon)\right]  & \text{at}~ x=-\varepsilon, \\
        -\frac{N_1^+}{p}\exp(-B_1\varepsilon)~\text{sech}^{A_1/\alpha}(\alpha\varepsilon)\sin(p\varepsilon)\left[Pp-Q\cot(p\varepsilon)\right]  & \text{at}~ x=\varepsilon.
    \end{cases}\nonumber\\
    &&\text{where},\nonumber\\
    &&P=-(A_1+A)\tanh(\alpha x)-(B_1+B)\quad\text{and}\nonumber\\
    &&Q=-B_1\left(B_1+B\right)-A_1(A_1+A)+(A_1+\alpha)(A_1+A)~\text{sech}^2(\alpha\varepsilon)\nonumber\\
    &&\qquad-\left\{B_1(A_1+A)+A_1(B_1+B)\right\}\tanh(\alpha\varepsilon).\label{E033}
\end{eqnarray}
Again, there are two distinct scenarios:
\begin{itemize}
    \item[{\bf Case 1}:] $N_1^-=-N_1^+=N_1$. As $D=0$,
    \begin{equation}
        Pp\tan{p\varepsilon}=Q,\label{E034}
    \end{equation}
    yielding,
    \begin{equation}
        C=N_1\exp(-B_1\varepsilon)~\text{sech}^{A_1/\alpha}(\alpha\varepsilon)\cos(p\varepsilon)\left(P+\frac{Q^2}{Pp^2}\right).\label{E035}
    \end{equation}
    As before, it is a parity-even eigenstate peaking at the origin with no minimum.
    \item[{\bf Case 2}:] With $N_1^-=N_1^+=N_1$, since $C=0$,
    \begin{equation}
        Pp\cot{p\varepsilon}=-Q,\label{E036}
    \end{equation}
    leading to,
    \begin{equation}
        D=-N_1\exp(-B_1\varepsilon)~\text{sech}^{A_1/\alpha}(\alpha\varepsilon)\sin(p\varepsilon)\left(P+\frac{Q^2}{Pp^2}\right).\label{E037}
    \end{equation}
    As a result, the combined state is again odd and vanishes at the origin as before.
\end{itemize}

\paragraph*{} In the DDF limit ($\varepsilon\to 0$, $p\to\infty$, $p\varepsilon\to 0$, $p^2\varepsilon\to\Omega/2$), Eq. \ref{E034} imposes the constraint,
\begin{equation}
    \boxed{\frac{\Omega}{2}=B_1-\alpha\frac{A_1+A}{B_1+B}},\label{E038A1}
\end{equation}
implying,
\begin{equation}
    C\to -(B_1+B)N_1,\label{E039}
\end{equation}
matching with the value of $\varphi_{I,III}(x)$ at $x=0$ from Eqs. \ref{E032}. Thus, the eigenstate in Case 1 survives. As for Case 2, under the same limit, Eq. \ref{E036} leads to the unacceptable conclusion of $Pp\to 0$ since $P=-(B_1+B)$ is finite. Further, $D\to 0$ from Eq. \ref{E037}. Therefore, the odd-parity case again does not survive.

\paragraph*{}As for the derivative discontinuity condition, for Case 1, 
\begin{equation}
    \left.\frac{d\psi_-^1}{dx}\right\vert_{x\to 0^-}^{x\to 0^+}=2N_1\left[\alpha(A_1+A)-B_1(B_1+B)\right].\label{E040}
\end{equation}
Then, from Eqs. \ref{E038A1} and \ref{E039},
\begin{equation}
    \lim_{\varepsilon\to 0}\int_{-\varepsilon}^\varepsilon dx V_-(x)\psi_1^-(x)=-2\Omega C=-2N_1\left[\alpha(A_1+A)-B_1(B_1+B)\right],\label{E041}
\end{equation}
and thereby the discontinuity condition is satisfied.

\subsection{2nd excited state}\label{A2}
The boundary conditions at $x=\pm\varepsilon$ on the wavefunction of Eq. \ref{E042} lead to,
\begin{eqnarray}
    &&C=
    \begin{cases}
        \frac{N_2^-}{p}\exp(-B_2\varepsilon)~\text{sech}^{A_2/\alpha}(\alpha\varepsilon)\left[Rp\cos(p\varepsilon)+S\sin(p\varepsilon)\right]  & \text{at}~ x=-\varepsilon, \\
        \frac{N_2^+}{p}\exp(-B_2\varepsilon)~\text{sech}^{A_2/\alpha}(\alpha\varepsilon)\left[Rp\cos(p\varepsilon)+S\sin(p\varepsilon)\right]  & \text{at}~ x=\varepsilon,
    \end{cases}\nonumber\\
    &&D=\begin{cases}
        -\frac{N_2^-}{p}\exp(-B_2\varepsilon)~\text{sech}^{A_2/\alpha}(\alpha\varepsilon)\left[Rp\sin(p\varepsilon)-S\cos(p\varepsilon)\right]  & \text{at}~ x=-\varepsilon, \\
        \frac{N_2^+}{p}\exp(-B_2\varepsilon)~\text{sech}^{A_2/\alpha}(\alpha\varepsilon)\left[Rp\sin(p\varepsilon)-S\cos(p\varepsilon)\right]  & \text{at}~ x=\varepsilon.
    \end{cases}\nonumber\\
    &&\text{where},\nonumber\\
    &&R=C_1-C_2~\text{sech}^2(\alpha\varepsilon)+ C_3\tanh(\alpha\varepsilon)\quad\text{and}\nonumber\\
    &&S=(A_2C_3+B_2C_1)-\left\{(\alpha+A_2)C_3+B_2C_2\right\}~\text{sech}^2(\alpha\varepsilon)\nonumber\\
    &&\qquad-(2\alpha+A_2)C_2~\text{sech}^2(\alpha\varepsilon)\tanh(\alpha\varepsilon)+(A_2C_1+B_2C_3)\tanh(\alpha\varepsilon).\label{E043}
\end{eqnarray}
For this system, the two possibilities are:
\begin{itemize}
    \item[{\bf Case 1}:] $N_2^-=N^+_2=N_2$ with $D=0$ leads to,
    \begin{equation}
        Rp\tan(p\varepsilon)=S.\label{E044}
    \end{equation}
    As a result,
    \begin{equation}
        C=N_2\exp(-B_2\varepsilon)~\text{sech}^{A_2/\alpha}(\alpha\varepsilon)\cos(p\varepsilon)\left(R+\frac{S^2}{Rp^2}\right),\label{E045}
    \end{equation}
    which depicts a smooth, parity-even eigenstate.
    \item[{\bf Case 2}:] $N^-_2=-N^+_2=N_2$ with $C=0$ leads to,
    \begin{equation}
        Rp=-S\tan(p\varepsilon).\label{E046}
    \end{equation}
    Subsequently,
    \begin{equation}
        D=-N_2\exp(-B_2\varepsilon)~\text{sech}^{A_2/\alpha}(\alpha\varepsilon)\sin(p\varepsilon)\left(R+\frac{S^2}{Rp^2}\right),\label{E047}
    \end{equation}
    which depicts a smooth, but parity-odd eigenstate.
\end{itemize}

\paragraph*{} In the DDF limit, for Case 1, Eq. \ref{E044} leads to the constraint,
\begin{equation}
   \boxed{\frac{\Omega}{2}=B_2-\alpha\frac{C_3}{C_1-C_2}=B_2-\alpha\frac{(A_2+A_1)(B_2+B)+(A_2+A)(B_2+B_1)}{(B_2+B_1)(B_2+B)-\alpha(A_2+A_1)}},\label{E048}
\end{equation}
along with,
\begin{equation}
    C=N_2(C_2-C_1)=N_2\left[(B_2+B_1)(B_2+B)-\alpha(A_2+A_1)\right].\label{E049}
\end{equation}
It can be checked that the derivative discontinuity condition is also satisfied. Case 2 is again unacceptable with $S\to-\infty$ and $D\to 0$.

\section{Eigenstates for the singular Rosen-Morse super-partner $V_+(x)$}\label{B}
\subsection{Ground state}\label{B1}
Imposing the boundary conditions at $x=\pm\varepsilon$ on the wavefunction of Eq. \ref{E054} leads to,
\begin{eqnarray}
    &&C=
    \begin{cases}
        \frac{M^-_0}{2p}\exp\left[(p-B_1)\varepsilon\right]~\text{sech}^{A_1/\alpha}(\alpha\varepsilon)(Tp+U) & \text{at}~ x=-\varepsilon, \\
        \frac{M^+_0}{2p}\exp\left[-(p+B_1)\varepsilon\right]~\text{sech}^{A_1/\alpha}(\alpha\varepsilon)(Tp-U)  & \text{at}~ x=\varepsilon,
    \end{cases}\nonumber\\
    &&D=\begin{cases}
        \frac{M^-_0}{2p}\exp\left[-(p+B_1)\varepsilon\right]~\text{sech}^{A_1/\alpha}(\alpha\varepsilon)(Tp-U)  & \text{at}~ x=-\varepsilon, \\
        \frac{M^+_0}{2p}\exp\left[(p-B_1)\varepsilon\right]~\text{sech}^{A_1/\alpha}(\alpha\varepsilon)(Tp+U)  & \text{at}~ x=\varepsilon.
    \end{cases}\nonumber\\
    &&\text{where},\nonumber\\
    &&T=-D_1+D_2~\text{sech}^2(\alpha\varepsilon)\quad\text{and}\nonumber\\
    &&U=-B_1D_1+B_1D_2~\text{sech}^2(\alpha\varepsilon)+(2\alpha+A_1)D_2~\text{sech}^2(\alpha\varepsilon)\tanh(\alpha\varepsilon)\nonumber\\
    &&\qquad+A_1D_1\tanh(\alpha\varepsilon).\label{E055}
\end{eqnarray}
Unlike the case of $V_-$, the above sets of expressions for $C$ and $D$ do not contradict. Their combination now leads to a general parametric constraint,
\begin{equation}
    \exp(4p\varepsilon)=\left(\frac{Tp-U}{Tp+U}\right)^2.\label{E056}
\end{equation}
The two roots of the above equation yield two distinct cases,
\begin{equation}
    \tanh(p\varepsilon)=
    \begin{cases}
        -\frac{U}{Tp} & :{\bf\text{ Case 1}}, \\
        -\frac{Tp}{U} & :{\bf\text{ Case 2}}.
    \end{cases}\label{E057}
\end{equation}
\begin{itemize}
    \item[{\bf Case 1}:] Substituting from Eq. \ref{E057} into Eqs. \ref{E055} and then equating the two expressions for $C$ (or $D$) one gets $M_0^-=M_0^+=M_0$. This subsequently leads to,
    \begin{equation}
        C=D=\frac{M_0}{2}T\exp(-B_1\varepsilon)~\text{sech}^{A_1/\alpha}(\alpha\varepsilon)~\text{sech}(p\varepsilon).\label{E058}
    \end{equation}
    Clearly, the eigenfunction is even across the origin with a peak at $x=0$.
    \item[{\bf Case 2}:] Here, we have $M_0^-=-M_0^+=M_0$. Consequently,
     \begin{equation}
        C=-D=-\frac{M_0}{2}T\exp(-B_1\varepsilon)~\text{sech}^{A_1/\alpha}(\alpha\varepsilon)~\text{cosech}(p\varepsilon).\label{E059}
    \end{equation}
    As a result, we get a parity-odd eigenfunction.
\end{itemize}

\paragraph*{} In the DDF limit, Case 1 alludes to the constraint condition $\boxed{\Omega/2=B_1}$, whereas,
\begin{equation}
    C=D=\frac{M_0}{2}\left[\alpha(A_1+A)-B_1^2+B^2\right].\label{E060}
\end{equation}
From Eqs. \ref{E054}, this yields $C+D=\varphi_{I,III}(x=0)$ as needed. On the other hand, for {\bf Case 2}, the constraint reduces to $T=0$, which is not an acceptable situation in general, along with $C=-D=0$. This is indeed a discontinuous wavefunction at the origin and needs to be rejected.

\subsection{1st excited state}\label{B2}
On imposing the boundary conditions on the wavefunction in Eq. \ref{E061},
\begin{eqnarray}
    &&C=
    \begin{cases}
        \frac{M^-_1}{2p}\exp\left[(p-B_2)\varepsilon\right]~\text{sech}^{A_2/\alpha}(\alpha\varepsilon)(Vp+W) & \text{at}~ x=-\varepsilon, \\
        \frac{M^+_1}{2p}\exp\left[-(p+B_2)\varepsilon\right]~\text{sech}^{A_2/\alpha}(\alpha\varepsilon)(-Vp+W)  & \text{at}~ x=\varepsilon,
    \end{cases}\nonumber\\
    &&D=\begin{cases}
        \frac{M^-_1}{2p}\exp\left[-(p+B_2)\varepsilon\right]~\text{sech}^{A_2/\alpha}(\alpha\varepsilon)(Vp-W)  & \text{at}~ x=-\varepsilon, \\
        -\frac{M^+_1}{2p}\exp\left[(p-B_2)\varepsilon\right]~\text{sech}^{A_2/\alpha}(\alpha\varepsilon)(Vp+W)  & \text{at}~ x=\varepsilon.
    \end{cases}\nonumber\\
    &&\text{where},\nonumber\\
    &&V=E_1-E_2~\text{sech}^2(\alpha\varepsilon)+E_3\tanh(\alpha\varepsilon)\quad\text{and}\nonumber\\
    &&W=(E_1B_2+E_3A_2)+\left\{(\alpha+A_2)E_3+E_2B_2\right\}~\text{sech}^2(\alpha\varepsilon)+(A_2-2\alpha)E_2~\text{sech}^2(\alpha\varepsilon)\tanh(\alpha\varepsilon)\nonumber\\
    &&\qquad+(E_1A_2+E_3B_2)\tanh(\alpha\varepsilon).\label{E062}
\end{eqnarray}
Similarly to the ground eigenstate, the above equation leads to two possible situations:
\begin{equation}
    \tanh(p\varepsilon)=
    \begin{cases}
        -\frac{W}{Vp} & :{\bf\text{ Case 1}}, \\
        -\frac{Vp}{W} & :{\bf\text{ Case 2}}.
    \end{cases}\label{E063}
\end{equation}
\begin{itemize}
    \item[{\bf Case 1}:] Substituting from Eq. \ref{E063} into Eqs. \ref{E062} and then equating the two expressions for $C$ (or $D$) one gets $M_1^-=-M_1^+=M_1$. This subsequently leads to,
    \begin{equation}
        C=D=\frac{M_1}{2}\exp(-B_2\varepsilon)~\text{sech}^{A_2/\alpha}(\alpha\varepsilon)~\text{sech}(p\varepsilon)V.\label{E064}
    \end{equation}
    The eigenfunction is even across the origin with a peak at $x=0$ since $\varphi_{I,III}$ have opposite signs.
    \item[{\bf Case 2}:] In this case, $M_1^-=M_1^+=M_1$. Thus,
     \begin{equation}
        C=-D=-\frac{M_1}{2}\exp(-B_2\varepsilon)~\text{sech}^{A_2/\alpha}(\alpha\varepsilon)~\text{cosech}(p\varepsilon)V.\label{E065}
    \end{equation}
    As a result, the eigenfunction is a odd-parity one.
\end{itemize}

\paragraph*{} In the DDF limit, the constraint condition in {\bf Case 1} reduces to,
\begin{equation}
    \boxed{\frac{\Omega}{2}=-\frac{W}{V}=\alpha\frac{(A_2+A_1)\left(A_2^2+B_2^2-A^2-B^2\right)}{(B_2+B_1)\left(B_2^2-B^2\right)-4\alpha(AB+A_1B_2)}-B_2},\label{E066}
\end{equation}
whereas,
\begin{equation}
    C=D=\frac{M_1}{2}V=\frac{M_1}{2}\left[(B_2+B_1)\left(B_2^2-B^2\right)-4\alpha(AB+A_1B_2)\right].\label{E067}
\end{equation}
Although the continuity is maintained, the maximum of $\varphi_{I,III}$ is not at $x=0$, leading to a dip there. As for Case 2, the constraint demands $V=0$ which is not acceptable. Further, from Eq. \ref{E065}, there is a break in the eigenstate at $x=0$ since $C=-D=0$ and thus this possibility is rejected.

\section{Ground state of Calogero system with DDF.}\label{D}
The regularized distribution of the potential in Eq. \ref{EIS01} is,
\begin{equation}
    V_\pm(x)=\begin{cases}
        \left(Ax+\frac{C}{x-\alpha}+B\right)^2\pm\left(A-\frac{C}{(x-\alpha)^2}\right)~~ \text{with}~W(x)=Ax+\frac{C}{x-\alpha}+B & \text{for}~x>\varepsilon\\
        \pm V_0 & \text{for}~-\varepsilon\leq x\leq\varepsilon\\
\left(Ax+\frac{C}{x+\alpha}-B\right)^2\pm\left(A-\frac{C}{(x+\alpha)^2}\right)~~ \text{with}~W(x)=Ax+\frac{C}{x+\alpha}-B & \text{for}~x<-\varepsilon
\end{cases},\label{EIS02}
\end{equation}
The superpotential(s) in Eq. \ref{EIS01} then yield a ground state distribution,
\begin{equation}
    \psi_0^-(x)=\begin{cases}
        \varphi_I(x)=N_0^-\left\vert x+\alpha\right\vert^{-C}\exp\left[-\frac{y_-^2}{2A}\right] & \text{for}~x<-\varepsilon\\
\varphi_{II}(x)=E\cos(px)+F\sin(px) & \text{for}~-\varepsilon\leq x\leq\varepsilon\\
\varphi_{III}(x)=N_0^+\left\vert x-\alpha\right\vert^{-C}\exp\left[-\frac{y_+^2}{2A}\right] & \text{for}~x>\varepsilon
    \end{cases},\label{EIS03}
\end{equation}
wherein $y_\pm=Ax\pm B$ and $p=\sqrt{V_0+E}\in\mathbb{R}$. Subsequent imposition of the boundary conditions at $x=\pm\varepsilon$ leads to,
\begin{eqnarray}
   && pE=\begin{cases}
        N^-_0\left\vert\varepsilon-\alpha\right\vert^{-C}\exp\left(-\frac{\theta^2}{2A}\right)\left[p\cos(p\varepsilon)+\left(\frac{C}{\varepsilon-\alpha}+\theta\right)\sin(p\varepsilon)\right], & x=-\varepsilon\\
        N_0^+\left\vert\varepsilon-\alpha\right\vert^{-C}\exp\left(-\frac{\theta^2}{2A}\right)\left[p\cos(p\varepsilon)+\left(\frac{C}{\varepsilon-\alpha}+\theta\right)\sin(p\varepsilon)\right], & x=\varepsilon
    \end{cases},\nonumber\\
    && pF=\begin{cases}
        -N_0^-\left\vert\varepsilon-\alpha\right\vert^{-C}\exp\left(-\frac{\theta^2}{2A}\right)\left[p\sin(p\varepsilon)-\left(\frac{C}{\varepsilon-\alpha}+\theta\right)\cos(p\varepsilon)\right], & x=-\varepsilon\\
        N_0^+\left\vert\varepsilon-\alpha\right\vert^{-C}\exp\left(-\frac{\theta^2}{2A}\right)\left[p\sin(p\varepsilon)-\left(\frac{C}{\varepsilon-\alpha}+\theta\right)\cos(p\varepsilon)\right], & x=\varepsilon
    \end{cases}.\label{EIS04}
\end{eqnarray}
The two possible scenarios are,
\begin{itemize}
    \item[{\bf Case 1}:] $N_0^-=N_0^+=N_0$ with $F=0$ implying,
    \begin{equation}
        p\tan(p\varepsilon)=\frac{C}{\varepsilon-\alpha}+\theta.\label{EIS05}
    \end{equation}
   This leads to,
    \begin{equation}
        E=N_0\left\vert\varepsilon-\alpha\right\vert^{-C}\exp\left(-\frac{\theta^2}{2A}\right)\cos(p\varepsilon)\left[1+\frac{\left(\frac{C}{\varepsilon-\alpha}+\theta\right)^2}{p^2}\right],\label{EIS06}
    \end{equation}
    forming an even eigenstate.
    \item[{\bf Case 2}:]  $N_0^-=-N_0^+=N_0$ with $E=0$ leads to,
    \begin{equation}
        p\cot(p\varepsilon)=-\frac{C}{\varepsilon-\alpha}-\theta.\label{EIS07}
    \end{equation}
    Subsequently,
    \begin{equation}
        F=-N_0\left\vert\varepsilon-\alpha\right\vert^{-C}\exp\left(-\frac{\theta^2}{2A}\right)\sin(p\varepsilon)\left[1+\frac{\left(\frac{C}{\varepsilon-\alpha}+\theta\right)^2}{p^2}\right].\label{EIS08}
    \end{equation}
    and teh corresponding eigenstate is odd.
\end{itemize}

\paragraph*{}In the DDF limit Case 1 corresponds to the constraint,
\begin{equation}
    \boxed{\Omega=-\frac{C}{\alpha}+B},\label{EIS09}
\end{equation}
along with the amplitude at $x=0$,
\begin{equation}
    E=N_0\left\vert\alpha\right\vert^{-C}\exp\left(-\frac{B^2}{2A}\right)=\varphi_{I,III}(x=0).\label{EIS10}
\end{equation}
Clearly, it is singular for $C>0$ once the regularization is removed ($\alpha=0$). On the other hand, Case 2 reduces to an unacceptable result of $C/\alpha-B=\infty$ even without the limit $\alpha\to 0$, with $F=0$.

\bibliographystyle{plain}

    \bibliography{Ref} 

\end{document}